\documentclass{aa} 
\usepackage{graphicx}
\usepackage{nameref}
\usepackage[varg]{txfonts}

\begin{document} 
	
	\title{Modeling effects of starspots on stellar magnetic cycles}
	
	
	\author{Zebin Zhang
		\inst{1}\fnmsep\inst{2}
		 \and
		Jie Jiang\inst{1}\fnmsep\inst{2}
		\and
		Leonid Kitchatinov\inst{3}\fnmsep\inst{4}
	}
	
	\institute{School of Space and Environment, Beihang University, Beijing, China\\
		\email{jiejiang@buaa.edu.cn}
		\and
		Key Laboratory of Space Environment monitoring and Information Processing of MIIT, Beijing, China
		\and
		Institute of Solar-Terrestrial Physics, Irkutsk, Russia
		\and
		Pulkovo Astronomical Observatory, St. Petersburg, Russia		
	}
	
	\date{Received xx, 2023; accepted xx, 2023}
	
	
	\abstract
	{Observations show that faster-rotating stars tend to have stronger magnetic activity and shorter magnetic cycles. The cyclical magnetic activity of the Sun and stars is believed to be driven by the dynamo process. The success of the Babcock-Leighton (BL) dynamo in understanding the solar cycle suggests an important role that starspots could play in stellar magnetic cycles.}
	{We aim at extending the BL mechanism to solar-mass stars with various rotation rates and explore the effects of emergence properties of starspots in latitudes and tilt angles on stellar magnetic cycles.}
	{We adopt a kinematic BL-type dynamo model operating in the bulk of the convection zone. The profiles of the large-scale flow fields are from the mean-field hydrodynamical model for various rotators. The BL source term in the model is constructed based on the rotation dependence of starspots emergence. That is, faster rotators have starspots at higher latitudes with larger tilt angles.}
	{Faster rotators have poloidal flux appearing closer to about $\pm55^\circ$ latitudes, where the toroidal field generation efficiency is the strongest because of the strongest latitudinal differential rotation there. It takes a shorter time for faster rotators to transport the surface poloidal field from their emergence latitude to the $\pm 55^\circ$ latitudes of efficient $\Omega$-effect thus shortening their magnetic cycles. The faster rotators operate in a more supercritical regime due to a stronger BL $\alpha$-effect relating to the tilt angles, which leads to stronger saturated magnetic fields and a coupling of the poloidal field between two hemispheres more difficult. Thus the magnetic field parity shifts from the hemispherically asymmetric mixed mode to quadrupole, and further to dipole when a star spins down.}
	{The emergence of starspots plays an essential role in the large-scale stellar dynamo.}
	
	
	\maketitle

\section{Introduction} 
The solar large-scale magnetic fields show the quasi-11-yr cyclic variations manifested by the sunspots \citep{Hathaway2015}. Starspots and magnetic cycles are ubiquitous among cool stars \citep{Strassmeier2009, Saikia2018}. The progress in solar magnetism paves the way for the understanding of stellar activity \citep{Brun2015}. 

The cyclical magnetic activity of stars is usually studied through the chromospheric emission, e.g., Ca II H$\&$K emission \citep{Wilson1978} or photospheric brightness variations \citep[e.g.,][]{Reinhold2017,Montet2017}. These studies show a general $P_{rot}$--$P_{cyc}$ relation, that is the magnetic cycle period $P_{cyc}$ tends to be longer for stars having longer rotation period $P_{rot}$ \citep{Noyes1984b}. Some studies show two branches of this relation, active and inactive ones corresponding to high and low activity, respectively \citep{Saar1999,BohmVitense2007,Wright2011}, while the existence of the active branch is still controversial \citep{Saikia2018}. Past efforts on stellar activity also reveal the dependence of magnetic activity amplitude on the rotation period, hereafter called the $P_{rot}$--$A_{cyc}$ relation. That is, younger and faster rotating stars tend to have higher magnetic activity amplitude $A_{cyc}$. When rotators are fast enough, magnetic activity amplitude tends to saturate at some level \citep{Hempelmann1996,Gudel2004, Wright2016, Zhang2020}.

The $P_{rot}$--$P_{cyc}$ relation of stellar magnetism provides an important observational test for dynamo models. In kinematic $\alpha$-$\Omega$ mean-field dynamo, the decreasing trend of the cycle period along the rotation period is a general property, because the cycle period is determined by the dynamo number related to the rotation period \citep{Tobias1998}. 
During the past decades, the flux transport dynamo \citep[FTD,][]{Wang1991,Durney1995,Choudhuri1995} works as the paradigm in understanding the solar cycle. In the framework of the FTD, the cycle period is controlled by the rate of meridional flow \citep{Dikpati1999,Jouve2007}. If extending the FTD model to stars, the $P_{rot}$--$P_{cyc}$ relation requires an increase of meridional flow as the rotation rate increases \citep{Nandy2004}. But magnetohydrodynamic (MHD) simulations show that the strength of meridional flow decreases as stars rotate faster \citep{Brown2008,Augustson2012}.
Thus the faster rotators host longer magnetic cycles \citep{Jouve2010,Karak2014}, which is contrary to what is observed. \cite{Kitchatinov2022} suggests that the effective temperature is the essential parameter in understanding stellar magnetic cycles. Hotter stars sustain shorter cycles and they rotate faster on average. 
\cite{Brun2017b} introduce muti-cell meridional flows to reconcile the discrepancy. 
\cite{DoCao2011} consider the effect of magnetic turbulent pumping, and found the magnetic cycle shortens when the pumping effect becomes stronger with the increase of rotation rate. 
\cite{Pipin2016} include the dynamical quenching of magnetic buoyancy and magnetic helicity and \cite{Hazra2019} introduce near-surface pumping and assume the faster rotators have a stronger BL source term. They both reproduce the relation between rotation period and magnetic cycle close to observations. 
MHD simulations report a weak relation between the rotation period and magnetic cycle with a negative slope \citep{Warnecke2018b,Strugarek2017,Brun2022}, while \cite{Guerrero2019} find a trend of the increase of the magnetic cycle with the rotation period under some conditions.

The $P_{rot}$--$A_{cyc}$ relation is another observational test for dynamo models. Faster rotators tend to form starspots with larger tilt angles, which could explain the $P_{rot}$--$A_{cyc}$ relation in the framework of the BL mechanism. Sunspots are formed by the buoyant rise of subsurface toroidal flux \citep{Parker1955b}. The Coriolis force acts on the rising toroidal flux, leading to tilt angles of sunspot groups \citep{Fan2021}. Starspots are expected to be formed in a similar process as sunspots. The Coriolis force increases for faster rotators and thus could lead to a larger tilt angle \citep{Caligari1995,Isik2018}. The tilt angles correspond to the strength of the BL source term, so faster rotators sustain a stronger BL source term. Based on the assumption that faster rotators sustain a stronger BL source term due to larger tilt angles, the $P_{cyc}$--$A_{cyc}$ relation was reproduced by \cite{Karak2014,Hazra2019}. Furthermore, \cite{Kitchatinov2015} considered the saturation of the BL mechanism for the tilt angles approaching $\pi /2$ and reproduced the magnetic activity saturation for very fast rotators. Some MHD simulations have also been carried out to understand the scaling law between stellar magnetic activity and rotation period \citep{Augustson2019,Brun2022}.

Solar observations show that stronger cycles tend to have smaller tilt angles \citep{DasiEspuig2010,Jiao2021}, which influence the contribution of sunspots to the polar fields. The polar fields at cycle minimum are believed to be the source of the toroidal field of a new cycle, which determines the strength of the subsequent solar cycle \citep{Jaramillo2013,Cameron2015,Jiang2018}. So the cycle-dependent tilt angles of sunspot emergence works as a nonlinear mechanism to modulate the amplitude of the solar cycle, which is called tilt quenching.
Observations also show that stronger cycles tend to have sunspot emergence at higher latitudes \citep{Li2003, Solanki2008,Jiang2011}. The higher latitudes the sunspots emerge, the weaker their contribution to build up the polar fields \citep{Jiang2014,Petrovay_2020}. So the cycle-dependent latitudes of sunspot emergence is another nonlinear feedback for the solar cycle, which is called latitude quenching. Systematic studies on the both forms of quenching have demonstrated that they play a crucial role in modulating the amplitude of the solar magnetic cycle \citep{Jiang2020, Karak2020,Talafha2022}. This raises an interesting question of how the starspots' emergence latitudes and tilt angles influence stellar magnetic cycles.

Sunspots appear only at latitudes lower than about 40$^\circ$. But for starspots, the emergence latitudes could be distributed over the whole stellar disk \citep{Strassmeier2009}. Observations also show faster rotators tend to have starspots at higher latitudes and even at polar regions \citep{Vogt1983,Strassmeier1991}.  
For the first time \cite{Schuessler1992} suggested an origin of the polar starspots, which result from a dominance of the Coriolis force in the dynamics of magnetic flux tubes emergence. \cite{Isik2018} calculate the emergence latitudes and tilt angles of starspots for various rotators, then apply the surface flux transport model to understand the evolution of the large-scale field at stellar surfaces. They found that faster rotators with starspots at higher latitudes reverse the polarity of the polar field in a shorter time, which implies faster rotators have shorter magnetic cycles. This work also inspires us to explore the effects of emergence properties of starspots on stellar magnetic cycles in the BL-type dynamo framework. 

\cite{Zhang2022a} recently develop a new generation of the BL-type dynamo model operating in the bulk of the convection zone, rather than in the tachocline. The model establishes a global dipolar fields connecting two solar poles. The configuration of the near surface poloidal field is consistent with observations and differs from most other available dynamo models. The poloidal field is sheared by latitudinal differential rotation and generate the strong toroidal field in the bulk of the convection zone as the source of sunspots. The tachocline is not important in their model, which is supported by stellar observations and MHD simulations that the dynamo process is more likely to operate in the bulk of the convection zone \citep{Nelson2013,Yadav2015,Wright2016}.

This work aims at extending the solar dynamo model of \cite{Zhang2022a} to stars with a solar mass ($M_\odot$), ranging the rotation periods from 10 to 25 days, and explore the effect of starspots emergence properties on magnetic cycles, and further test if the $P_{rot}$--$P_{cyc}$ and $P_{rot}$--$A_{cyc}$ relations can be reproduced due to starspots' properties. On the other hand, the tests of stellar observations could also verify the solar dynamo model. We will quantify the dependence of starspots' emergence properties in tilt angles and latitudes on the rotation rate of stars based on observations. 

The paper is organized as follows. The BL-type dynamo model is described in Sect. 2. We determine the critical dynamo number in Sect. 3.1.1. The effect of the emergence latitude of starspots on the magnetic cycle is explored in the linear regime in Sect. 3.1.2. The $P_{cyc}$--$A_{cyc}$ relation is studied in the nonlinear regime in Sect. 3.2. We summarize our results in Sect. 4.

\section{MODEL} \label{sec:model}
In this study, we adopt the kinematic dynamo model developed by \cite{Zhang2022a}, who deal with the evolution of axisymmetric large-scale magnetic field, $\textbf{B}(r,\theta,t)=B(r,\theta,t) \hat{\textbf{e}}_\phi+
\nabla\times
\left[A(r,\theta,t)\hat{\textbf{e}}_\phi\right]$, with prescribed flow profiles in the standard spherical polar coordinates ($r,\theta,\phi$). The dynamo equations are expressed as 
\begin{equation} \label{eq:dynamo_A}
	\frac{\partial A}{\partial t}+\frac{1}{s}[(\textbf{u}_{p}+\gamma_{r}\hat{\textbf{e}}_{r})\cdot\nabla](sA)
	=\eta\left(\nabla^{2}-\frac{1}{s^{2}}\right)A+S_{BL}, 
\end{equation}

\begin{equation}  \label{eq:dynamo_B}
	\begin{split}
	\frac{\partial B}{\partial t}+\frac{1}{r}\left[\frac{\partial(u_{r}+\gamma_{r})rB}
	{\partial r}+\frac{\partial(u_{\theta}B)}{\partial\theta}\right]=\eta\left(\nabla^{2}-\frac{1}
	{s^2}\right)B+\\
	s(\textbf{B}_{p}\cdot\nabla\Omega)+\frac{1}{r}\frac{d\eta}
	{dr}\frac{\partial(rB)}{\partial r}, 
	\end{split}
\end{equation} 
where $\eta$ and $\gamma_r$ represent the turbulent diffusivity and the radial pumping, respectively. Differential rotation and meridional flow are represented by $\Omega(r,\theta)$ and $\textbf{u}_{p}(r,\theta) = u_r(r,\theta)\hat{\textbf{e}}_{r}+u_{\theta}(r,\theta)\hat{\textbf{e}}_{\theta}$, respectively. The parameters will be presented in the next subsections.

\begin{figure*}[h]
	\centering
	\includegraphics[width=13cm]{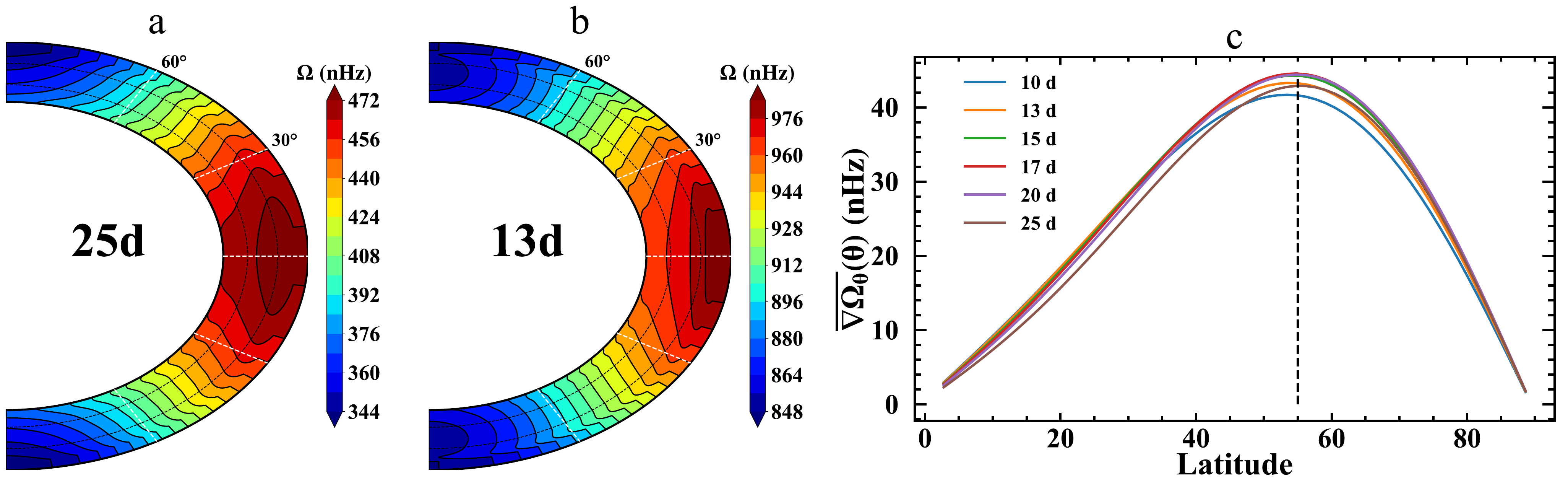}	
	\caption{Differential rotations of solar-mass stars. Panels (a) and (b) are the profiles of differential rotation for stars with rotation periods of 25 and 13 days, respectively. Panel (c) shows the latitudinal distribution of the radius-averaged latitudinal shear $\overline{\nabla\Omega_{\theta}}(\theta)$ for the different rotators we study. The vertical dashed line denotes 55$^\circ$ latitude.}
	\label{Figure_Diff}
\end{figure*}

\begin{figure*}[h]
	\centering
	\includegraphics[width=13cm]{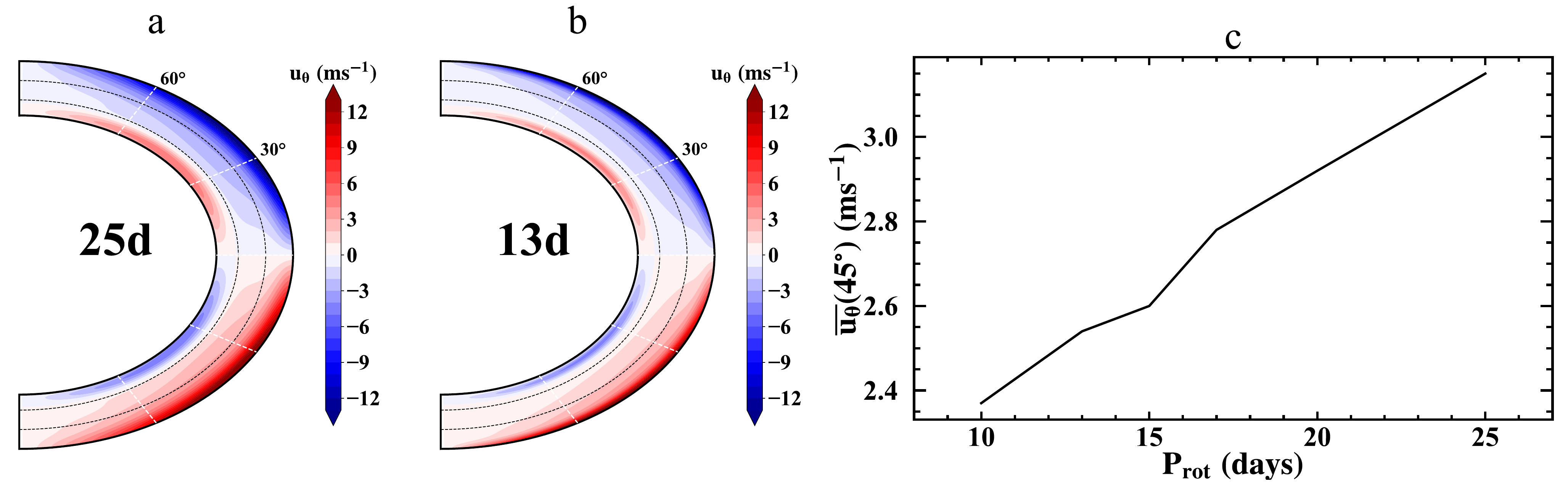}	
	\caption{Meridional flows of solar-mass stars. Panels (a) and (b) are the latitudinal component of the meridional flow patterns for the stars with rotation periods of 25 and 13 days, respectively. Panel (c) shows the dependence of the averaged return flow at 45$^\circ$,  $\overline{u_{\theta}}(45^\circ)$, on the stellar rotation period.}
	\label{Figure_MF}
\end{figure*}

\subsection{Differential rotation and meridional flow} 
The large-scale flows in 1$M_\odot$ stars rotating with different rates are specified with the method developed by \cite{Kitchatinov1999}. The mean-field model based on this method \citep{Kitchatinov2011,Kitchatinov2012} predicts the dependence of differential rotation on the spectral type and rotation rate in agreement with observations by \cite{Barnes2005} and \cite{Balona2016}. Here, the version of the differential rotation model by \cite{Kitchatinov2017b} adapted for using in dynamo simulations is applied.
The results have been adopted to understand properties of the stellar magnetic field \citep[e.g.][]{Hazra2014, Kitchatinov2022}. Observations indicate that faster rotators tend to have stronger torsional oscillations \citep{CollierCameron2002} and more irregular stellar cycles \citep{Baliunas1995, Saikia2018}. 
Hence we restrict our computations to slowly rotating 1$M_\odot$ stars with the rotation periods from 10 to 25 days.

Figures \ref{Figure_Diff} (a) and (b) show the profiles of differential rotation for stars with sidereal rotation periods of 25 and 13 days, respectively. The former one corresponds to the solar case and is close to helioseismic results \citep{Schou1998}. With the increase of the rotation rate, the differential rotation changes toward a cylinder-shaped pattern. It is the latitudinal shear in the bulk of the convection zone that winds the poloidal field for the dynamo model we use. Figure \ref{Figure_Diff} (c) shows the latitudinal distribution of the latitudinal shear, $\overline{\nabla\Omega_{\theta}}(\theta)=2\pi \int_{0.72R_\odot}^{R_\odot}  \nabla \Omega_\theta (r, \theta) \,dr$. With the increase of the rotation rate, the strength of latitudinal shear increases slightly, which is consistent with observations \citep{Barnes2005}. Similar to the Sun, all stars have the strongest latitudinal shear at about $\pm$55$^{\circ}$ latitude, where the generation efficiency of the toroidal field is the strongest and corresponds to the seat of stellar dynamo processes \citep{Zhang2022a}.

Figures \ref{Figure_MF} (a) and (b) show the $\theta$ component of the meridional flow, $u_{\theta}$, for stars with rotation periods of 25 and 13 days, respectively. For all rotators, the flow has a single-cell structure in each hemisphere, and is poleward near the surface and equatorward at the bottom of the convection zone. 
Figure \ref{Figure_MF} (c) shows the dependence of the equatorward return flow, $\overline{u_{\theta}}(45^{\circ})$, on the rotation period. The radius-averaged return flow, $\overline{u_{\theta}}(\theta)$, is defined as $\overline{u_{\theta}}(\theta)= \int_{0.72R_\odot}^{R_c}  u_\theta(r, \theta) \,dr/\int_{0.72R_\odot}^{R_c}\,dr$, where $R_c$ is the depth that the equatorward flow starts for different rotators. The strength of $\overline{u_{\theta}}(45^{\circ})$ decreases from 3.15 m s$^{-1}$ to 2.4 m s$^{-1}$ with the decrease of the rotation period from 25 days to 10 days. This trend is consistent with 3-D convective simulations \citep{Brown2008,Augustson2012}. The slower return flow for faster rotators is the reason why the FTD dynamo models failed to reproduce observational P$_{rot}$--P$_{cyc}$ relation \citep{Karak2014}.

\subsection{Turbulent pumping and diffusivity}
A near-surface radial pumping $\gamma_r$ is included in our model aiming at matching the observed large-scale magnetic field evolution at the solar surface \citep{Cameron2012,Jiang2013}. We adopt its profile as
\begin{equation}
	\gamma_r(r)=-\frac{\gamma_{0}}{2}\left[1+\rm erf\left(\frac{r-r_s}
	{d_s}\right)\right],\label{eq8}
\end{equation}
where $r_s=0.9R_{\odot}$ and $d_s=0.01R_\odot$ making sure that the pumping is confined near surface and smoothly decreases to zero at 0.88$R_{\odot}$.
A large enough $\gamma_{0}$ near surface is used just to prevent the diffusive escape of magnetic flux through the solar surface. Its amplitude and penetration depth are free parameters and has no effect on the cycle period of the dynamo model. Here we set $\gamma_{0} = 20$ m s$^{-1}$ for all rotators.

Turbulent diffusivity $\eta$ within the stellar convection zone is still poorly constrained. Assuming that the turbulent convection dominates $\eta$ in the convection zone and it gets significantly reduced through the overshot layer. We adopt the following diffusivity profile 
\begin{equation}
	\eta=\frac{\eta_{cz}}{2}\left[1+\rm erf\left(\frac{r-r_c}{d_c}\right)\right]+\frac{\eta_{s}}{2}\left[1+\rm erf\left(\frac{r-r_s}{d_s}\right)\right],\label{eq9}
\end{equation}
where $r_c=0.7R_\odot$ and $d_c=0.03R_\odot$ correspond to the center and thickness of the overshot layer.
The turbulent diffusivity in the bulk of the convection zone $\eta_{cz}$ is taken as 3.7$\times10^{7}$ m$^{2}$ s$^{-1}$, which is closer to the value estimated by the mixing-length theory than most values used by FTD models \citep[see Figure 1 of][]{Munoz2011}. The diffusivity near the surface $\eta_{s}$ is taken as the supergranular diffusivity $3.0\times10^{8}$ m$^{2}$ s$^{-1}$, which is within the range of observational studies summarized in Table 1 of \cite{Schrijver1996}.

\subsection{Babcock-Leighton source term}
The BL-type source term shown in Eq.(\ref{eq:dynamo_A}) is the core of the model setup to achieve our main objective of exploring the influence of starspots' emergence characteristics on stellar magnetic cycles.
It is defined as 
\begin{equation}
	S_{BL}(r,\theta,t)=\alpha(r,\theta) \bar{B}(\theta,t).\label{eq:source}
\end{equation}
The poloidal field is produced from the mean toroidal field $\bar{B}(\theta,t)$ in the bulk of the convection zone from $0.72R_\odot$ to $0.88R_\odot$, i.e., 
\begin{equation}
	\bar{B}(\theta,t)=\int_{0.72R_\odot}^{0.88R_\odot}B(r,\theta,t)
	rdr/\int_{0.72R_\odot}^{0.88R_\odot}rdr,\label{eq5}
\end{equation} 
where 0.72 $R_\odot$ and 0.88 $R_\odot$ corresponds to the inner boundary and penetration depth of the pumping, respectively.

The $\alpha$-effect term can be written as
\begin{equation}
	\alpha(r,\theta)=\frac{\alpha_{0}}{2}g(r)f(\theta).\label{eq_alpha_rThetha_0}
\end{equation}
The radial dependence of the $\alpha$-effect is to constrain the BL process just working near the surface based on the essence of the BL process. It is
rewritten in the form
\begin{equation}
	g(r)=\left[1+\rm erf\left(\frac{r-r_\alpha}{d_\alpha}\right)\right], 
	\label{eq_alpha_r}
\end{equation}
where $r_\alpha=0.95R_\odot$ and $d_\alpha=0.01R_\odot$.

\begin{figure}[h]
	\centering
	\includegraphics[width=9cm]{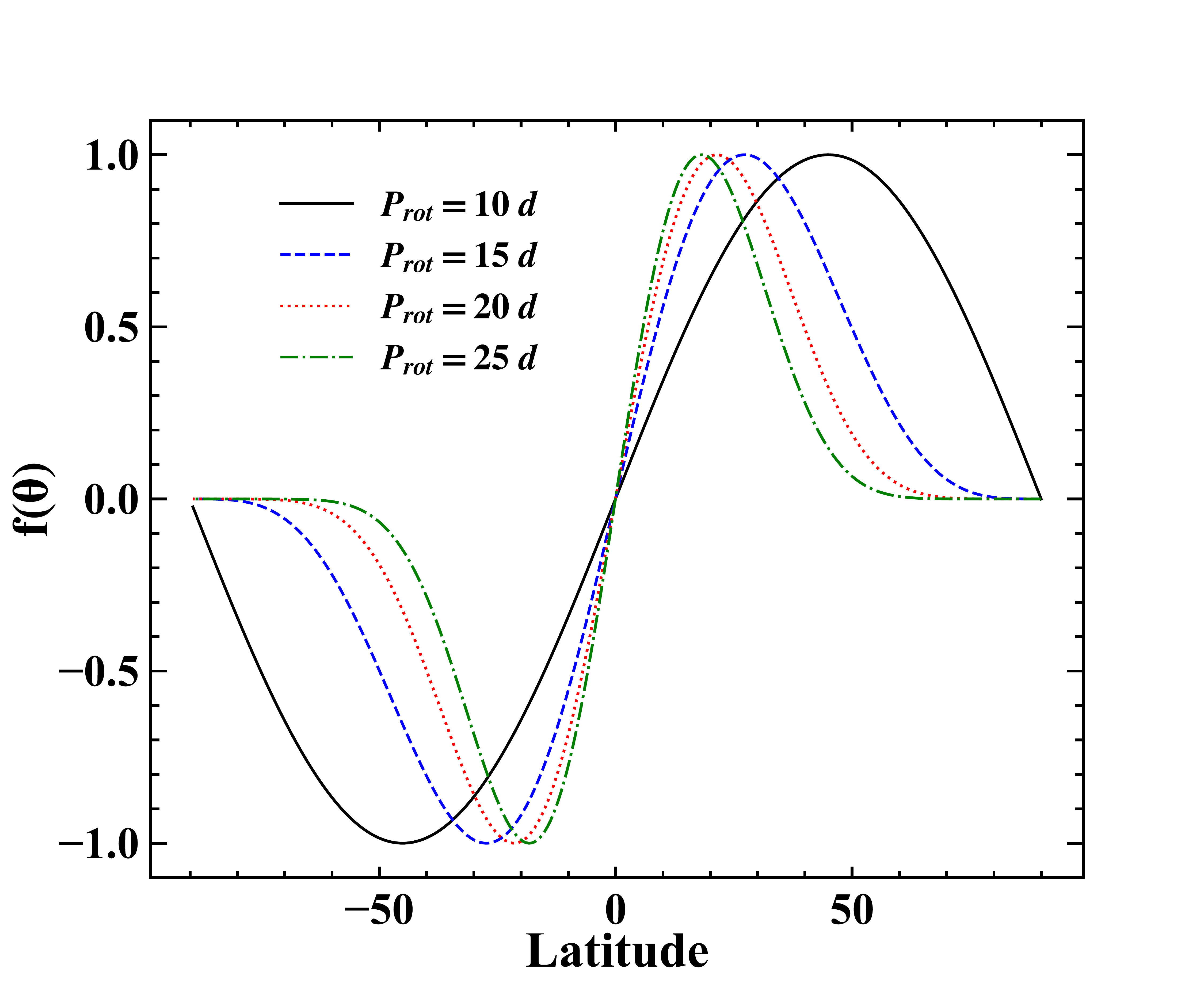}
	\caption{Latitude variation of the $\alpha$ profiles determined by Eq. (\ref{eq6}) for stars with different rotation periods.}
	\label{Figure1_alpha}
\end{figure}

To explore how the latitudes of starspots emergence influence magnetic cycles, we design the $f(\theta)$ term as 
\begin{equation}
	f(\theta)=\dfrac{\cos\theta \sin^{n}\theta}{Max[\cos\theta\sin^{n}\theta, \theta\in(0,\pi)]} ,\label{eq6}
\end{equation}
where $\cos\theta$ reflects the latitude dependence of the tilt angles caused by the Coriolis force and $Max[\cos\theta\sin^{n}\theta, \theta\in(0,\pi)]$ keeps the maximum value of Eq. (\ref{eq6}) the same for the various $n$ values. The factor $sin^{n}\theta$ reflects the dependence of the probability of toroidal flux emergence on latitude \citep{Cameron2017}. We suppose a linear relation between the rotation period $P_{rot}$ of stars and the $n$ value,  
\begin{equation}
	n = 1+(P_{rot}-10)*0.55, \label{eqn}
\end{equation}
so that it is $n = 1$ for $P_{rot}$ = 10 days and  $n = 9.25$ for $P_{rot}$ = 25 days. With the decrease of rotation rate, the $n$ value increases and constrains the emergence of starspots to lower latitudes.  Figure \ref{Figure1_alpha} shows latitudinal variations of Eq. (\ref{eq6}) for various rotators. As stars rotate faster, $n$ decreases and the maxima of the curves shift to higher latitudes. For the rotation period from 25 days to 10 days, the latitude of maximum $\alpha$ changes from 18$^\circ$ to 45$^\circ$. 

Note that Eq. (\ref{eqn}) is formulated to mimic the observation of faster rotators displaying starspots at higher latitudes, which are nearer to $\pm$55$^\circ$ latitudes. The trend is in agreement with the thin-flux-tube simulations by \cite{Isik2018}, who utilized a surface flux transport model to investigate stellar brightness variations for different levels of magnetic activity and rotation rates. To date, neither observations nor numerical simulations have been able to determine the exact dependence of starspots’ latitudes on $P_{rot}$. We will also consider a quadratic relationship between $n$ and $P_{rot}$, that is  $n = 9.25(P_{rot}/P_{sun})^2$, and independence of $n$ on the rotation period, that is $n=1$ in Eq. (\ref{eqn}) as a contrast, in Sect. \ref{sec_result_Linear}.

Another important parameter of the BL source term is its strength $\alpha_0$, dominated by tilt angles of starspots. The dynamo number, $\alpha_0 \Delta\Omega R^3/\eta^{2}$, is a key parameter in the dynamo process. It describes the efficiency of magnetic fields generation versus decay caused by diffusion. A self-sustained dynamo process occurs when the dynamo number exceeds a critical value. After given a specific profile of diffusivity, differential rotation and the source term of the poloidal field, the strength of the dynamo number is determined by the free parameter $\alpha_0$ in Eq. (\ref{eq_alpha_rThetha_0}). So that a non-decay dynamo process occurs when $\alpha_0$ is greater than or equal to its critical value $\alpha_c$. In most previous dynamo models, $\alpha_{0}$ was determined arbitrarily. \cite{Kitchatinov2017a} (hereafter KN17) suggest that $\alpha_{0}$ could be constrained based on observations.

KN17 assumes that the upper bound $P_{max}$ for the rotation period of solar-type stars \citep{vanSaders2016} corresponds to near marginal dynamo excitation. In other words, the faster-rotating stars tend to have a stronger $\alpha$-effect and operate in a more supercritical regime. For the BL mechanism, $\alpha_0$ is dominated by the tilt angles. The Coriolis force is stronger for faster rotators, thus leading to a larger tilt angle of starspots and then larger $\alpha_0$. The $\alpha_0-\alpha_c$ describes how supercritical the dynamo is. Motivated by KN17, we give $\alpha_0$ for a star with the rotation period $P_{rot}$ based on 
\begin{equation} \label{eq:alpha0_gyr}
	\alpha_0 = (1+m\frac{P_{max}}{P_{rot}})\alpha_c,
\end{equation}
where $m$ is a free parameter constraining the amount of supercriticality. We take $P_{max}$ as 28 days for the solar-mass stars we study.

The saturated unsigned toroidal flux within the convection zone is around $10^{23}$ Mx for the solar case \citep{Cameron2015}. The flux puts a constraint on the $m$ value, which is taken as $m = 0.2$. This entails that supercriticality varies from about 20\% to 56 \% for rotators with $P_{rot}$ from 25 to 10 days.

\subsection{Initial and boundary conditions} \label{sec:IC}
The configuration of large-scale magnetic fields in the Sun and stars can be classified as equatorially symmetric (quadrupolar) and anti-symmetric (dipolar) parity. The latter one is the dominant parity in the Sun, 
while some observations imply that fast rotators might have dominant quadrupolar fields \citep{Kochukhov2021}.
In this work, we start the simulations with two kinds of seed magnetic fields, whereas the toroidal field is set to be 
\begin{equation}
B(r, \theta)|_{t=0} = \sin(2\theta)\sin[\pi(r-0.72R_\odot)/0.28R_\odot]
\label{eq10}
\end{equation}
for dipolar parity, or 
\begin{equation}
B(r, \theta)|_{t=0} = \sin(\theta)\sin[\pi(r-0.72R_\odot)/0.28R_\odot]
\label{eq11}
\end{equation}
for quadrupolar parity. The poloidal field is set to be zero. 

The outer boundary condition satisfies the vertical field condition based on the constraint by \cite{Cameron2012}. Accordingly, we use $\partial (rA)/ \partial r = 0$, $B=0$ at $r = R_\odot$. The inner boundary matches a perfect conductor, which means that $A = 0$, $\partial(rB) / \partial r = 0$ at $r = 0.72R_\odot$. At poles, $A = B = 0$. The computational domain of our model is 0.72$R_\odot \leq r \leq $ $R_\odot, 0 \leq \theta \leq \pi$. Note again that our model does not include the tachocline. 

Our model is calculated using a code that utilizes the Crank-Nicolson scheme and an approximate factorization technique \citep{Houwen2001}. The code developed at Beihang university has been validated against the open-source Surya developed by A.R. Choudhuri and his colleagues \citep{Chatterjee2004} and dynamo benchmark \citep{Jouve2008}.

\section{Results}
\label{Sec: results}
For the onset of dynamo instability, controlling parameters should exceed a certain critical value. The $\alpha_0$ of Eq.(\ref{eq_alpha_rThetha_0}) is the only variable parameter in our model since other parameters involved in the dynamo number, i.e., differential rotation and turbulent diffusivity, are given. Here we will first look for $\alpha_c$ for the dipolar and quadrupolar solutions and analyze properties of different symmetric solutions for different stars in Sect. \ref{sec_result_Linear}. Then we will use the constrained $\alpha_0$ given by Eq. (\ref{eq:alpha0_gyr}) to analyze the dynamo behaviors of different stars in Sect. \ref{sec:nonlinearCase}. 

\subsection{Linear model} \label{sec_result_Linear}
\subsubsection{Critical $\alpha$-values, $\alpha_c$, for dipolar and quadrupolar modes}
 
With the time-independent large-scale flows and $\alpha$ independent of the magnetic field, dynamo Eqs. (\ref{eq:dynamo_A}) and (\ref{eq:dynamo_B}) are fully linear. Thus both $A$ and $B$ have exponential time dependence in the form of $e^{\lambda t}$, with $\lambda=\sigma+i \omega$. The real part $\sigma$ is a growth rate, and the imaginary part $\omega$ is an oscillation frequency satisfying $\omega=\frac{2\pi}{P_{cyc}}$. The solution with zero linear growth rate ($\sigma=0$) is the purely oscillatory one. The corresponding $\alpha_0$ is the critical value denoted as $\alpha_{c}$. Usually, $\alpha_c$ is derived as an eigenvalue problem of the linear system \citep[e.g.,][]{Jiang2005,Bonanno2002,Jiang2007}. Here we solve the dynamo equations as an initial value problem. With a given initial condition, we try different $\alpha_0$ values. The larger $\alpha_0$ value is, a faster grow rate $\sigma$ is.  The $\alpha_c$ is pinned down when $|\sigma| < 10^{-3}$ measured by the grow rate of the toroidal field integrated through the convection zone. In the linear system, the parity of fields is determined by the seed initial fields. For example, the seed fields with a pure dipolar parity could only excite a dipole-mode solution. So we use initial fields of dipolar (Eq. (\ref{eq10})) and quadrupolar (Eq. (\ref{eq11})) parity to find the critical values $\alpha_{c}^d$ and $\alpha_{c}^q$ for dipolar and quadrupolar parity, respectively. 

\begin{figure}[h]
	\centering
	\includegraphics[width=9.5cm]{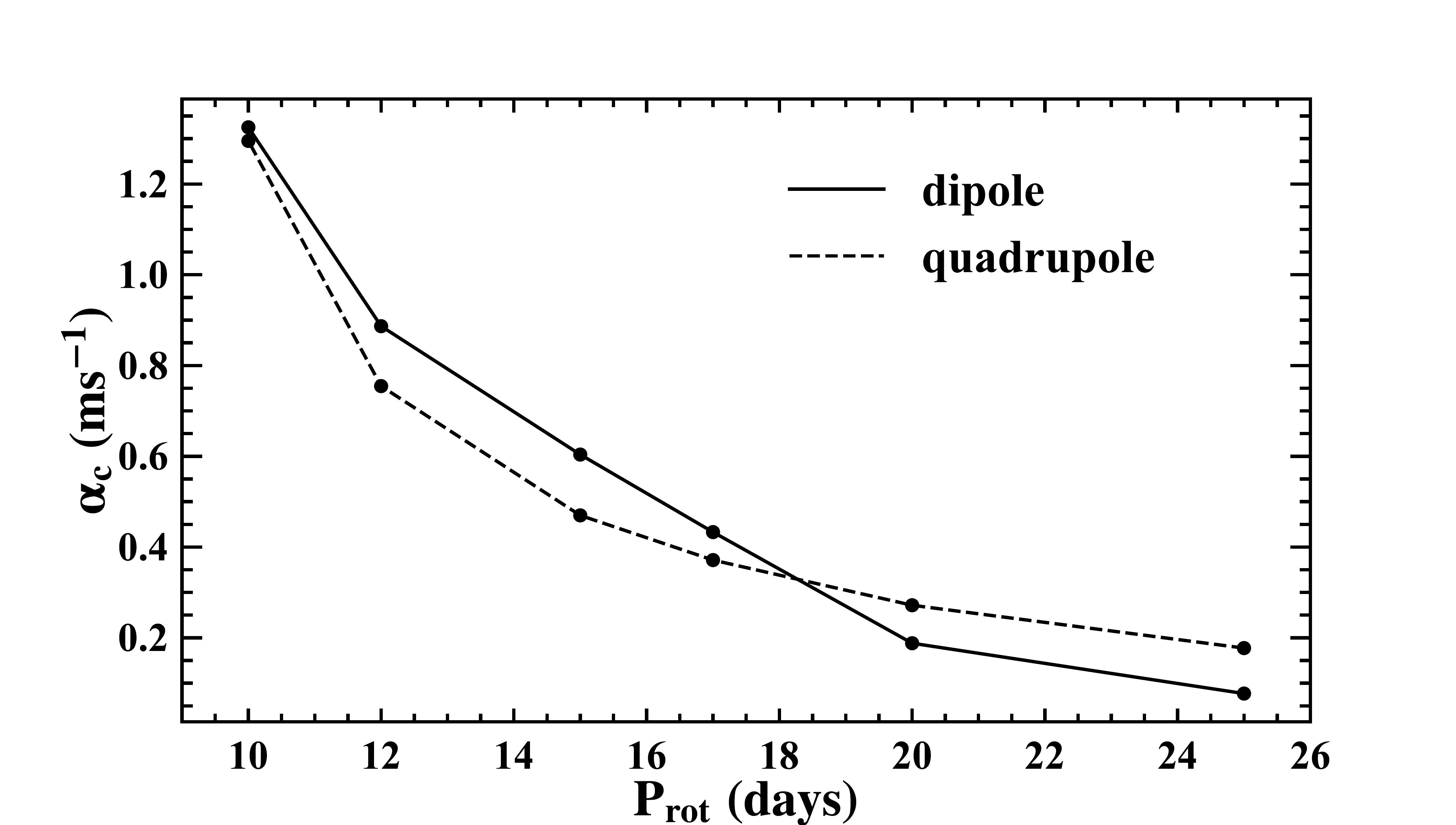}	
	\caption{Critical $\alpha$-value, $\alpha_c$, as a function of stellar rotation period ($P_{rot}$). The solid (dashed) curve represents the dipolar (quadrupolar) parity solution.}
	\label{Figure_cri}
\end{figure}

Figure \ref{Figure_cri} shows $\alpha_c$ as a function of the stellar rotation period. For the Sun (sidereal rotation period about 25 days), $\alpha_{c}^d$ and $\alpha_{c}^q$ are 0.08 m s$^{-1}$ and 0.18 m s$^{-1}$, respectively. The solution of dipolar parity has the smaller $\alpha_{c}$, which means that the dipolar parity is easier to excite. This is consistent with the observed solar magnetic field, which is dominated by the dipolar field. Incidentally, $\alpha_{c}$ for the kinetic helicity at the base of the solar convection zone is about 10 m s$^{-1}$ based on the estimation of \cite{Charbonneau2020}. Furthermore, Figure \ref{Figure_cri} shows that $\alpha_c$ increases with the rotation rate. The increase rate of the dipolar solutions (in the solid curve) differs from the quadrupolar one (in the dashed curve). The rotation period of 18 days marked a turning point. The dipolar (quadrupolar) solution prevails when the stellar rotation period is longer (shorter) than 18 days. Larger $\alpha_c$ indicates stronger diffusive annihilation of magnetic fields for a given system. The next subsection, especially Figure \ref{Figure_new1}, will show that slower rotators have a simpler configuration of the magnetic field corresponding to lower order multipoles, which entails weaker diffusive annihilation \citep{Wang2001}.

\subsubsection{$P_{rot}$ -- $P_{cyc}$ relation and parity property}
\begin{figure}[h]
	\centering
	\includegraphics[width=9.5cm]{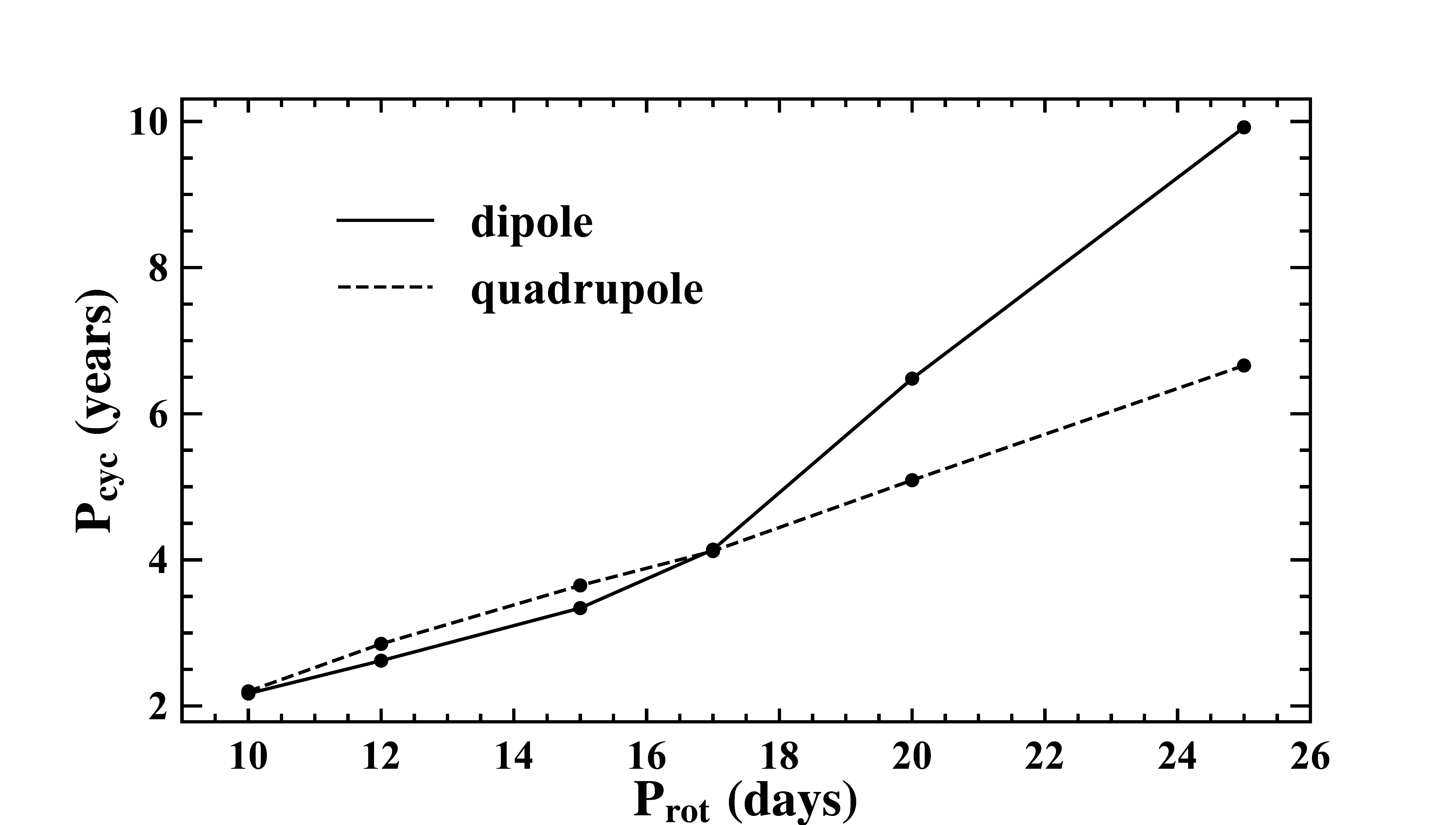}	
	\caption{Relation between rotation period ($P_{rot}$) and magnetic cycle ($P_{cyc}$). The solid (dashed) curve represents the dipolar (quadrupolar) parity solution.}
	\label{Figure_PP}
\end{figure}

Figure \ref{Figure_PP} gives the relation between the rotation period and the magnetic cycle near marginal dynamo
excitation. The solid (dashed) curve represents the dipolar (quadrupolar) solutions, both of which show that the magnetic cycle generally increases with the rotation period. 

To understand what dominates the magnetic cycles in our model, we investigate the dependence of $P_{cyc}$ on two parameters, the average speed of return meridional flow $\overline{u_{\theta}}$ and the critical number $\alpha_c$. It turns out that $P_{cyc}^{d}\propto \overline{u_{\theta}}^{2.4}$ and $P_{cyc}^{q}\propto \overline{u_{\theta}}^{2.0}$. When the meridional flow is faster, the magnetic cycle is longer. This indicates that our model is different from the FTD models, in which the relation between $P_{cyc}$ and $\overline{u_{\theta}}$ obeys $P_{cyc}^{d}\propto \overline{u_{\theta}}^{x}$, where $x=-0.89$ given by \cite{Dikpati1999} and $x=-0.696$ given by \cite{Karak2010}. 
On the other hand, it suggests $P_{cyc}^{d}\propto (\alpha_{c}^{d})^{-0.79}$ and $P_{cyc}^{q}\propto (\alpha_{c}^{q})^{-0.91}$. The magnetic cycle decreases with $\alpha_{c}$ and hence with the dynamo number. Faster rotators have a higher efficiency of dynamo process and then sustain a shorter magnetic cycle. In all simulations, we keep the pumping term the same. In models with turbulent pumping through the whole convection zone, the flux transport by pumping influences the cycle period \citep{Guerrero2008,Hazra2016}. The near-surface pumping introduced in our model only makes the surface part of the dynamo process consistent with observations \citep{Cameron2012, Jiang2013}. It does not affect the cycle period. 

\begin{figure*}[h]
	\centering
	\includegraphics[width=12cm]{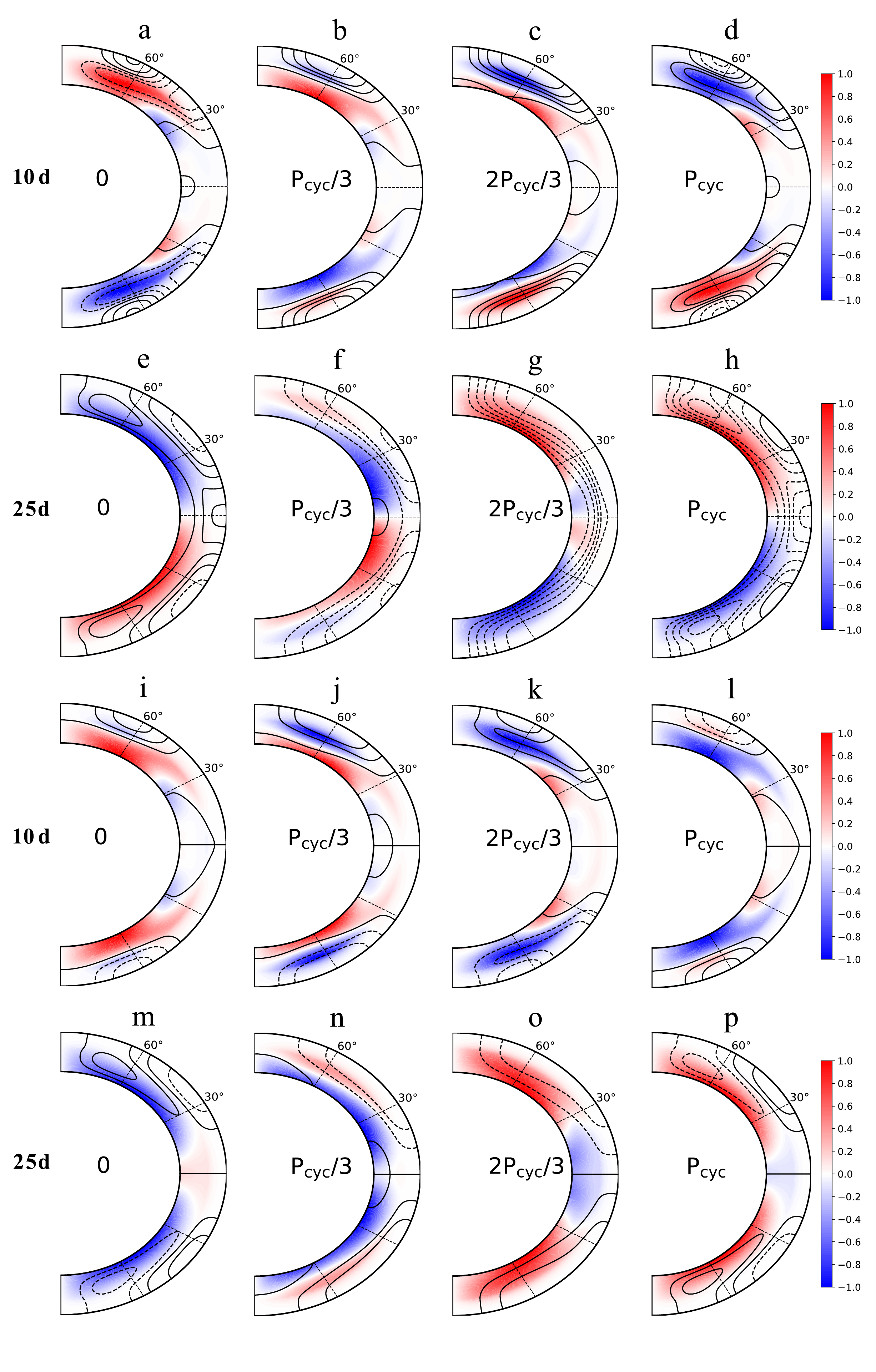}
	\caption{Snapshots of the magnetic field evolution during the interval of one third of magnetic cycle $P_{cyc}$/3. First (third) row represents rotation period of 10 days with dipolar (quadrupolar) fields. Second (forth) row represents rotation period of 25 days with dipolar (quadrupolar) fields. The strength of magnetic field is arbitrary in a linear regime, and it is normalized to a maximum of 1 here. The region in red (blue) represents toroidal fields inward (outward). The solid (dashed) lines represent the poloidal field clockwise (anticlockwise).}
	\label{Figure_new1}
\end{figure*}

To further explore why faster rotators have a shorter magnetic cycle period, we plot the time evolution of dipolar (quadrupolar) magnetic fields for $P_{rot}$ = 10 days in the first (third) row of Fig. \ref{Figure_new1}. Figure \ref{Figure_new1} (a) shows that the poloidal flux relating to starspots emergence first appears around $\pm55^\circ$ latitudes, where the latitudinal shear is the strongest. Therefore the newly generated poloidal flux could be sheared efficiently, thus generating the toroidal field for the subsequent cycle. The quick production of the toroidal field at high latitudes can be seen from Figs. \ref{Figure_new1} (b) - (c). Then the toroidal field gives rise to the poloidal field as illustrated by the dashed poloidal magnetic field lines in Fig. \ref{Figure_new1} (d). Thus a new magnetic cycle starts. The subsequent evolution of the poloidal field reverses the poloidal flux system of the old cycle. High-latitude starspots emergence shorten the dynamo process and hence the magnetic cycle. 

As a comparison, we show the corresponding results of the solar case with $P_{rot}$ = 25 days in the second row of Fig. \ref{Figure_new1}. Figure \ref{Figure_new1} (e) shows that the newly generated poloidal flux at surface first appears around $\pm$35$^\circ$ latitudes, which results from the setup of $n=9.25$ in Eq.(\ref{eq6}) to mimic the observation that sunspot emergence of each new cycle start from about $\pm$35$^\circ$. Then the poloidal flux is transported poleward, equatorward, and inward simultaneously. The surface turbulent diffusion leads to the equatorward migration, so that the net flux of one polarity resulting from the tilt angle diffuses across the equator. And this helps to finally establish the global dipolar field as shown in Fig. \ref{Figure_new1} (g). The poleward meridional flow along with the turbulent diffusion poleward transports the net flux of the other polarity to higher and higher latitudes. After a time interval $\Delta t_1$, they are transported to around the $\pm$55$^\circ$ latitudes, where the latitudinal shear is the strongest, and the poloidal field is transported to a deeper depth by the inward turbulent diffusion as well. Thus the toroidal field of a new cycle first appears around $\pm$55$^\circ$ (see Fig. \ref{Figure_new1} (f)). These toroidal fields could correspond to the ephemeral regions observed in the Sun \citep{Zhang2022a}. 
The lower latitudes have a weaker latitudinal rotational shear, and hence it takes more time to wind up the poloidal field to generate the strong enough toroidal field for sunspot emergence of a new cycle. During the wind-up process, the newly generated weak toroidal field is transported to lower latitudes due to the effect of the equatorward return flows, which entails further more time to be wound up to generate the strong enough toroidal field. So that the toroidal field of the new cycle is gradually built up at the lower latitudes as demonstrated by Figs. \ref{Figure_new1} (f)-(h). We denote the time delay caused by the latitude dependence of the latitudinal shear as $\Delta t_2$. Comparing with the case of $P_{rot}$ = 10 days, it takes more time, $\Delta t_1 + \Delta t_2$, to complete a dynamo loop for the case of $P_{rot}$ = 25 days, and hence it has longer cycle period.

\begin{figure}[h]
	\centering
	\includegraphics[width=9cm]{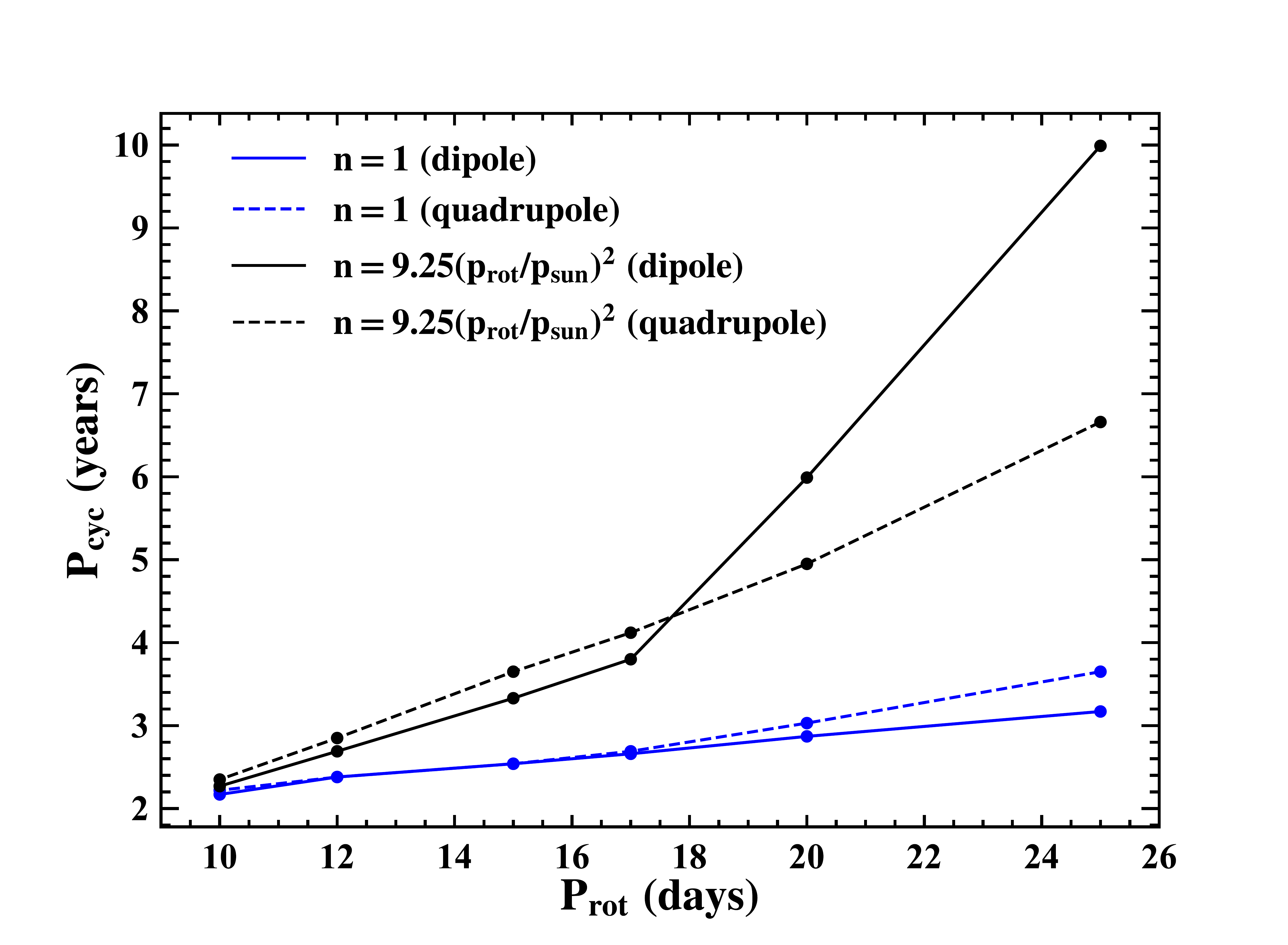}	
	\caption{Relation between rotation period ($P_{rot}$) and magnetic cycle ($P_{cyc}$), for cases of n = 1 (blue curve) and n = 9.25$(p_{rot}/p_{sun})^2$ (black curve). The solid (dashed) curve represents the dipolar (quadrupolar) parity solution.}
	\label{A1}
\end{figure}

In summary, the emergence property of starspots in latitudes plays a crucial role in creating the observed relation between $P_{rot}$ and $P_{cyc}$. 
Faster rotators have starspots in closer proximity to the $\pm55^\circ$ latitudes, which have the strongest latitudinal differential rotation and, therefore, the highest toroidal field generation efficiency. 
The strongest latitudinal differential rotation around the $\pm55^\circ$ latitudes has been confirmed on the Sun through helioseismology results.
Slower rotators with starspots at lower latitudes require extra time to transport and wind up the poloidal flux, resulting in longer cycle periods. As long as the faster rotators have starspots located closer to latitudes of $\pm55^\circ$, the observed $P_{rot}$ -- $P_{cyc}$ relation is expected to be reproduced. To verify this deduction, we utilize the equation $n = 9.25 (P_{rot}/P_{sun})^2$ to replace Eq. (\ref{eqn}). As $P_{rot}$ ranges from 25 days to 10 days, the latitude of maximum $\alpha$ changes from 18$^\circ$ to 40$^\circ$. The black curves in Figure \ref{A1} indicate the simulated relation between $P_{rot}$ and $P_{cyc}$ for dipolar (solid curve) and quadrupolar (dashed curve) solutions. They exhibit similar results as those in Figure \ref{Figure_PP}. On the other hand, the blue curves in Figure \ref{A1} show the results for $n=1$ in Eq. (\ref{eqn}), independent of $P_{rot}$. There is a slight increase in $P_{cyc}$ with the increase of $P_{rot}$. The increase is too small to account for the observed property. The increase is caused by variations in the meridional flow. Slow rotators have a greater return flow that can transport more toroidal flux towards lower latitudes. This leads to a decrease in the latitude of flux emergence and further a slight increase of the cycle period.

Figures \ref{Figure_cri} and  \ref{Figure_PP} have shown that for fast rotators ($P_{rot} <$ 18 days), the dipolar and quadrupolar modes have the similar cycle periods. And the quadrupolar mode has a smaller $\alpha_c$, hence is easier to excite. For slow rotators ($P_{rot} \geqslant$ 18 days), the dipolar mode has the longer cycle period and is easier to excite than the quadrupolar mode. The time evolution of the quadrupolar field for the cases of $P_{rot}$ = 10 days and $P_{rot}$ = 25 days is shown in the third and the fourth rows of Fig. \ref{Figure_new1}, respectively. 
For faster rotators, such as the case of $P_{rot} =$ 10 days, the surface poloidal field concentrated around middle-high latitudes. Most equatorward poloidal flux is canceled by the newly generated flux system of a new cycle, also for the toroidal flux. Thus the whole dynamo process operates locally and there are few coupling of magnetic fields between two hemispheres. The degree of the coupling of the poloidal field between two hemispheres determines the parity dominating in the dynamo process \citep{Chatterjee2004, Hotta2010}. 
For slower rotators, such as the case of $P_{rot} =$ 25 days, there are the across-equator cancellation of the toroidal field for the dipolar mode. These cancellations entail a longer time for the poloidal flux generation of the new cycle to reverse that of the old cycle. The global dipolar mode has the slowest decay in the absence of sources \citep{Cameron2010}. Hence it has a smaller $\alpha_c$ than the quadrupolar mode.

\subsection{$P_{rot}$ -- $A_{cyc}$ and $P_{rot}$ -- $P_{cyc}$ relations in the nonlinear regime} \label{sec:nonlinearCase}
We have investigated the dynamo process in the linear regime and found a dominant role of the starspots' emergence latitudes in determining magnetic cycle. To explore the relation between stars' rotation period $P_{rot}$ and amplitude of stellar activity $A_{cyc}$, the dynamo model working in the nonlinear regime is required. We adopt the same algebraic quenching term as that in \cite{Zhang2022a} as the nonlinear amplitude-limitation mechanism here.
The initial condition can be any arbitrary linear combinations of the dipolar and quadrupolar field presented by Eqs. (\ref{eq10}) and (\ref{eq11}), which do not affect the final magnetic field evolution. The strength of the BL source term $\alpha_0$ is prescribed by Eq. (\ref{eq:alpha0_gyr}) depending on both $\alpha_{c}$ and $P_{rot}$, where $\alpha_{c}$ is set as smaller of $\alpha^{d}_{c}$ and $\alpha^{q}_{c}$. Other ingredients of the dynamo model are the same as the one in the linear regime.

\begin{figure}[b]
	\centering
	\includegraphics[width=9cm]{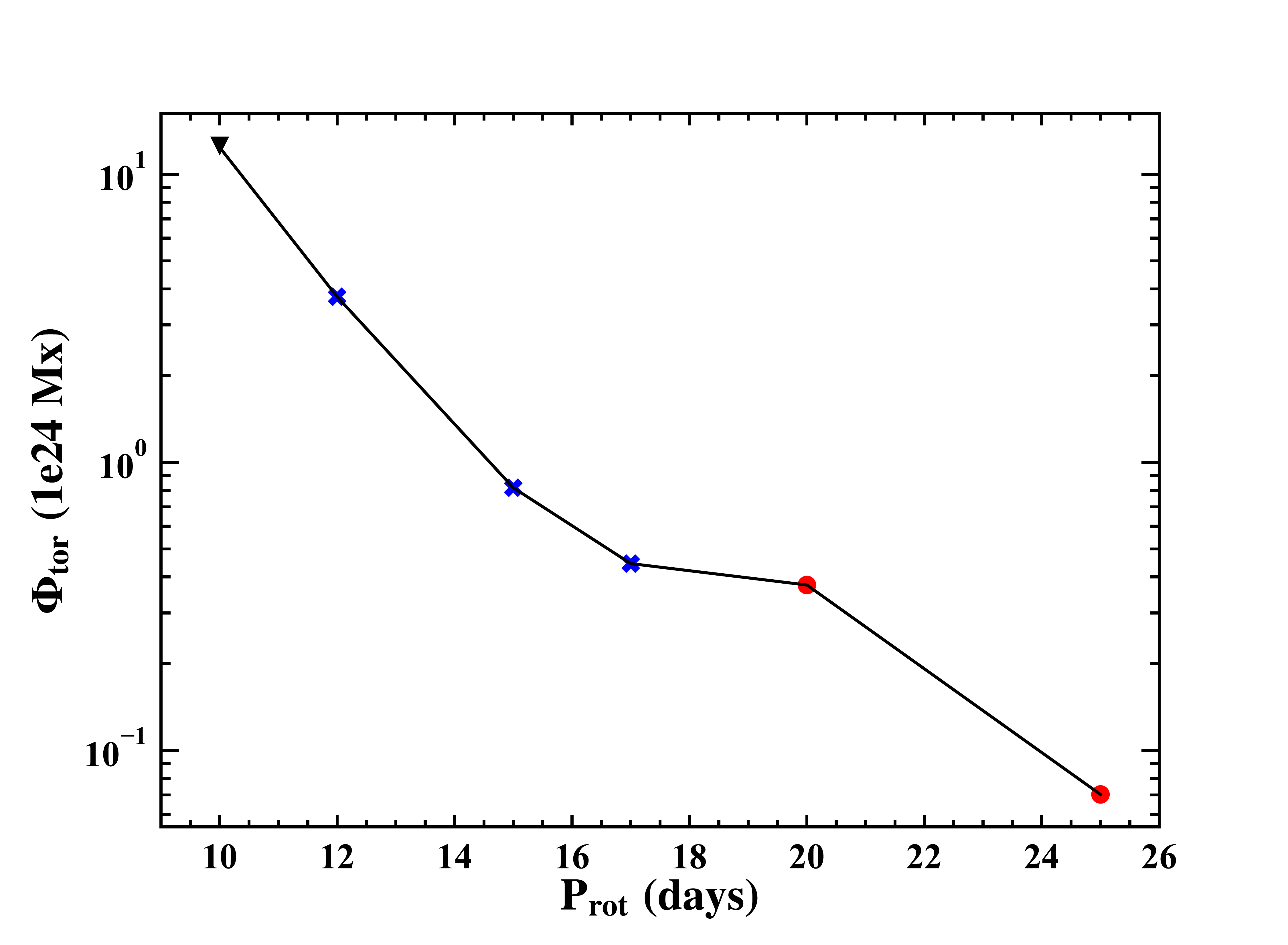}	
	\caption{Relation between rotation period ($P_{rot}$) and magnetic activity amplitude ($A_{cyc}$). The activity amplitude is measured by $\Phi_{tor}$. Red dots, blue forks, and black inverted triangle represent the saturated magnetic fields with a dipolar, quadrupolar, and mixed parity, respectively.}
	\label{Figure10_PA}
\end{figure}

In a dynamo model using algebraic quenching, the saturated magnetic field depends on the value $\alpha_0 - \alpha_{c}$. In this work, we adopt the hypothesis from KN17 to determine $\alpha_0$ based on $\alpha_{c}$ and $P_{rot}$. So an increase of magnetic activity amplitude $A_{cyc}$ with rotation rate $P_{rot}$ is expected, which is demonstrated by Fig. \ref{Figure10_PA}. Here magnetic activity amplitude $A_{cyc}$ is measured by cycle-averaged unsigned subsurface toroidal flux, $\Phi_{tor}$, 
\begin{equation}
	\Phi_{tor}=\int_{0}^{\pi} \int_{0.72R_\odot}^{R_\odot}  |B|r \,dr d\theta.
	\label{eq_Phi_tor}
\end{equation}
For the case of $P_{rot}$ = 25 days, $\Phi_{tor}$ is around 10$^{23}$ Mx, which is consistent with observations \citep{Cameron2015}. In contrast, \cite{Hazra2019} show a dip in their $P_{rot}$ -- $A_{cyc}$ relation around $P_{rot}$ = 15 days. The dip results from the different ways to deal with the strength of the BL source term $\alpha_{0}$. They did not estimate $\alpha_c$ for various rotators. So that the value of $\alpha_0 - \alpha_c$ is not consistent with the expected trend of monotonic increase with the decrease of $P_{rot}$ based on gyrochronology.

\begin{figure}[h]
	\centering
	\includegraphics[width=9cm]{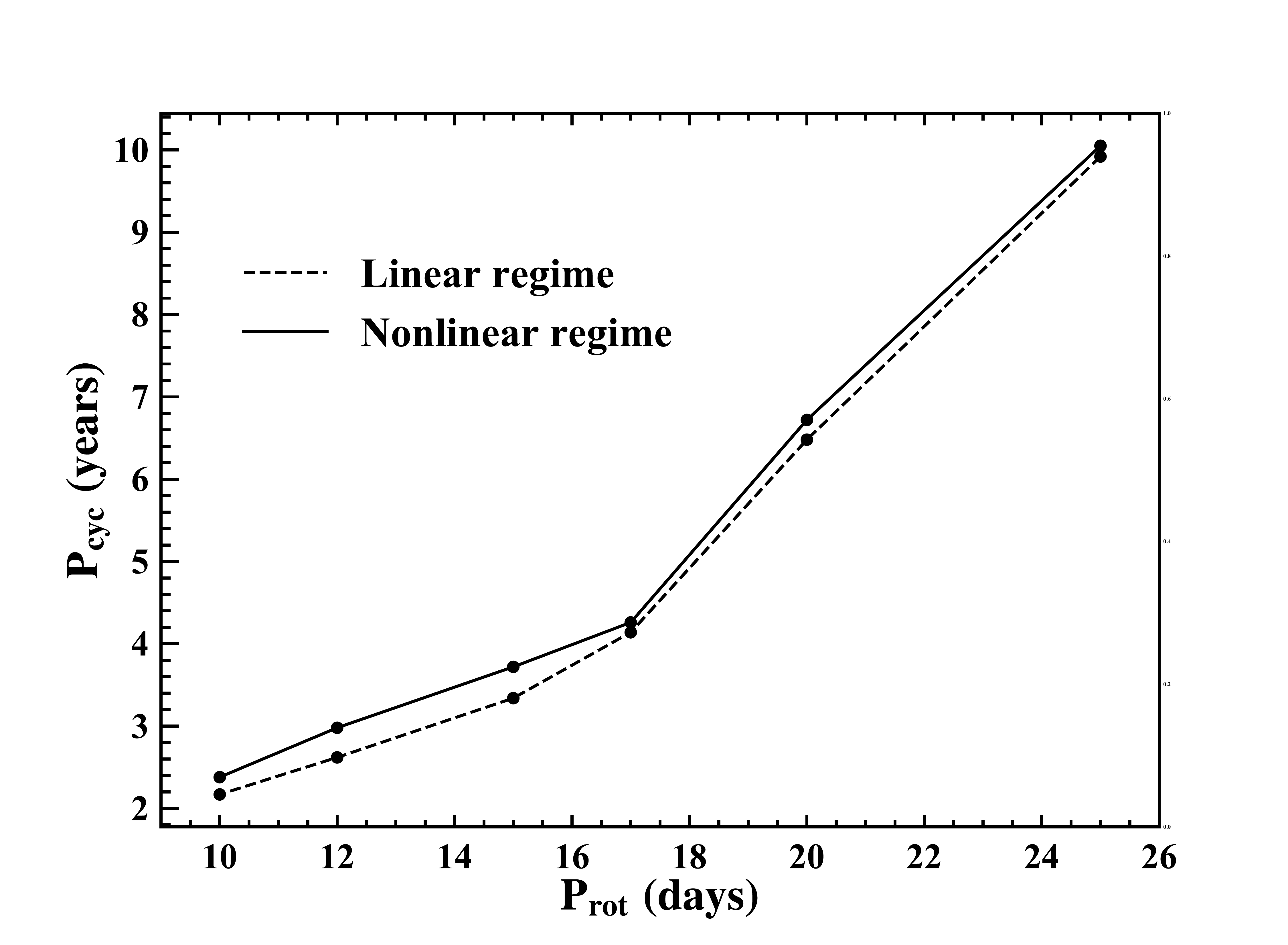}	
	\caption{Relation between rotation period ($P_{rot}$) and magnetic cycle ($P_{cyc}$). The solid (dashed) line represents the solution operating in the nonlinear regime (linear regime).}
	\label{Figure10_PP}
\end{figure}

\begin{figure*}[h]
	\centering
	\includegraphics[width=15cm]{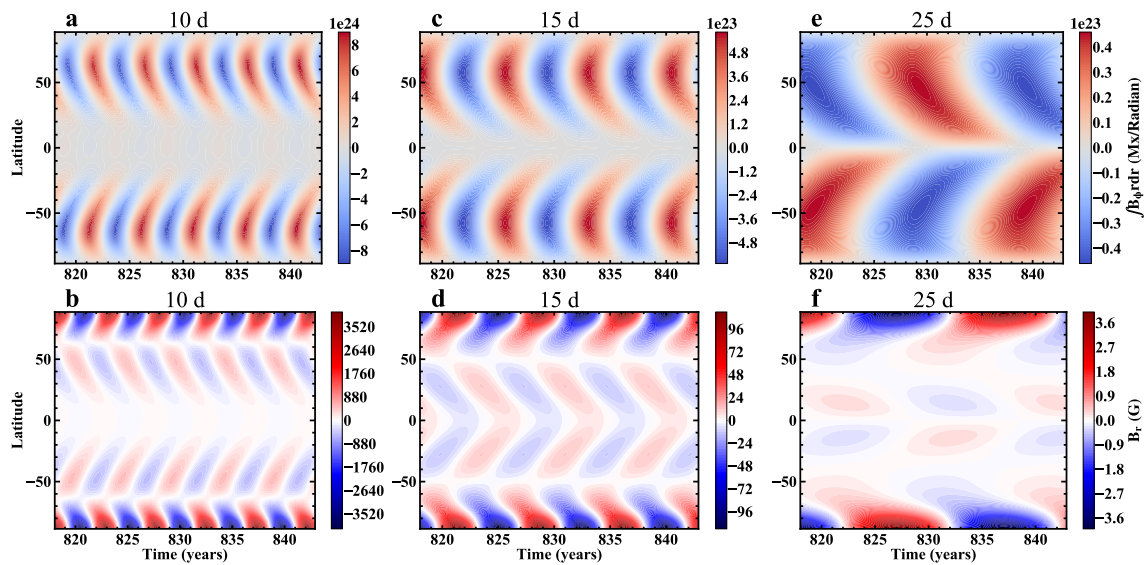}	
	\caption{Time-latitude diagrams of subsurface toroidal flux (top) and surface radial field (bottom) for stars with rotation periods of 10, 15 and 25 days.}
	\label{Figure:Nonlinear_Bphi_Br}
\end{figure*}

Figure \ref{Figure10_PP} shows the $P_{rot}$ -- $P_{cyc}$ relation in both the linear and nonlinear regimes. The magnetic cycles of the nonlinear solutions are slightly longer than that of the linear critical solutions, because algebraic quenching weakens the efficient dynamo number to its critical number as the toroidal field increases \citep{Noyes1984a, Tobias1998}. We fit the simulated $P_{rot}$ -- $P_{cyc}$ relation in the nonlinear regime and obtain $P_{cyc}$ $\propto$ $P_{rot}^{0.85}$, which is close to the observed inactive branch. 
As demonstrated in the last subsection, the rotation-dependent emergence latitude of starspots is the major reason for our model to reproduce the $P_{rot}$ -- $P_{cyc}$ relation, while \cite{Hazra2019} suggested that different profiles of the BL source term had no effect on the $P_{rot}$--$P_{cyc}$ relation. The two seemingly contradictory statements are caused by two different dynamo models used by the two papers. For \cite{Hazra2019}, the toroidal field is mainly generated by the radial shear in the tachocline, which has the strongest value near the poles. Thus although \cite{Hazra2019} adopted two profiles of the BL source term, the surface poloidal field distributions are similar for the two cases they used since the strong toroidal field near the poles dominates the surface poloidal field distributions.

Figure \ref{Figure_cri} has shown $\alpha_{c}^d < \alpha_{c}^q$ for the slowly rotators with $P_{rot}$ > 18 days and $\alpha_{c}^d > \alpha_{c}^q$ for the fast rotators with $P_{rot} < $ 18 days. These indicate that in the weakly nonlinear regime, slowly (fast) rotating stars host dipolar (quadrupolar) global fields. For the case of $P_{rot}$ = 10 days, $\alpha_0$ is 56\% over the critical value according to Eq. (\ref{eq:alpha0_gyr}). The global fields host a mixed parity because the strong super-criticality could lead to the appearance of hemispherically asymmetric mixed mode field  \citep{Jennings1991}.
Figure \ref{Figure:Nonlinear_Bphi_Br} shows the time-latitude diagram of subsurface toroidal flux (top panels) and surface radial field (bottom panels) for stars with rotation periods of 10, 15, and 25 days. 
For the case of $P_{rot}= $ 25 days, Figures \ref{Figure:Nonlinear_Bphi_Br} (e) and (f) well present properties of the solar cycle, such as the dipolar field, regular polar field reversals every about 11 years, and latitudinal migration of the toroidal field. With the decrease of the rotation period from 25 days to 15 days, the surface poloidal field appears at higher latitudes. The magnetic cycle period becomes shorter, activity amplitude is stronger, and the global field shifts to the quadrupolar parity. For the case of $P_{rot}= $ 10 days, Figures \ref{Figure:Nonlinear_Bphi_Br} (a) and (b) show the mixed-mode solution. Besides the increase of the toroidal flux with the increase of the rotation rate, the amplitude of the polar field also increases with the rotation rate. The trend is consistent with the surface flux transport results given by \cite{Schrijver2001, Isik2018}. The several thousand Gauss polar field could be responsible for the polar starspots in fast rotators.

\section{Conclusion and discussion}
We have extended the BL mechanism to solar-mass stars and explored the effect of emergence properties of starspots on stellar magnetic cycles. The rotation-dependent emergence latitude and tilt angle of starspots are introduced in a BL-type dynamo working in the bulk of the convection zone developed by \cite{Zhang2022a}. 
The model accounts well for the observed $P_{rot}$--$P_{cyc}$ and $P_{rot}$--$A_{cyc}$ relations about stellar magnetic activity. 

We are not the first to extend the BL mechanism developed based on solar observations to understand stellar magnetic cycles. Some efforts had been taken in this direction \citep[e.g.,][]{Jouve2010,Karak2014,Kitchatinov2015,Vashishth2023}. The past efforts mainly emphasized that the poloidal field results from the tilt angle of starspots and is generated near the stellar surface. The rotation-dependent tilt angle of starspots was considered by \cite{Hazra2019, Kitchatinov2022}. To the best of our knowledge, we are the first to include the rotation-dependent emergence latitude into the BL mechanism to understand stellar cycles.
We have also demonstrated that without the rotation-dependent emergence latitude, the resulting $P_{rot}$--$P_{cyc}$ relation is inconsistent with the observation.
The rotation-dependent emergence latitude of starspots is a property presented by observations of stellar magnetic activity. 
On the other hand, an essential role of the cycle-dependent emergence latitude of sunspots in modulating solar cycles, i.e., latitudinal quenching, was recognized recently \citep{Jiang2020, Karak2020}. The progress in the understanding of the solar cycle naturally extends to solar-type stars.

Actually some recent studies have focused on effects of the latitudinal distribution and tilt of starspots on stellar magnetic activity, e.g., brightness variations \citep{Isik2018, Nemec2023} and astrometric jitter \citep{Sowmya2021} based on surface flux transport models \citep{Baumann2004, Jiang2014b}. The flux emergence properties in latitude and tilt are prescribed by a separate model. These studies assume that different rotators have the same cycle period, i.e., 11 years, as the solar magnetic cycle period. The observed $P_{rot}$--$P_{cyc}$ relation was not included.

In our BL-type dynamo model, the emergence latitudes of starspots play a crucial role in reproducing the $P_{rot}$--$P_{cyc}$ relation. Faster rotators have flux emergence closer to $\pm55^\circ$ latitudes, where the latitudinal differential rotation and the toroidal field generation efficiency are the strongest. This entails a shorter cycle period. Slower rotators have flux emergence at lower latitudes. It takes some time for the surface poleward meridional flow and turbulent diffusion to transport the emergent poloidal field to around $\pm55^\circ$ latitudes. The poloidal flux is transported inwards and equatorward simultaneously. It takes further time for the interior poloidal field at lower latitudes to be wound up to generate a strong enough toroidal field because the lower latitudes have a weaker latitudinal shear. The extra time caused by the surface flux transport and the latitude dependence of the latitudinal shear in the interior leads to the longer cycle period $P_{cyc}$ for slower rotators.
There are no strict constraints on the relationship between the emergence latitude and rotation. As long as faster rotators have starspots closer to the $\pm55^\circ$ latitudes, the observed $P_{rot}$--$P_{cyc}$ relation can be reproduced. Without the rotation dependence of the starspots’ emergence latitudes, that is n = 1 in Eq. (\ref{eqn}), the resulting $P_{rot}$--$P_{cyc}$ relation is inconsistent with the observed one.
The latitude dependence of the latitudinal shear also has a large contribution to the equatorward migration of the toroidal field, i.e., the so-called butterfly diagram \citep{Zhang2022b}. 

The aforementioned explanations of the observed $P_{rot}$--$P_{cyc}$ relation based on our BL-type dynamo model discriminate our model developed by \cite{Zhang2022a} from the flux transport dynamo models \citep{Karak2014b}, in which the equatorward return flow dominates the cycle period and the equatorward migration of the toroidal field. We will analyze the detailed physical ingredients of \cite{Zhang2022a} accounting for the cycle period and butterfly diagram in a forthcoming paper.

In our BL-type dynamo models, the $55^\circ$ latitude is regarded as the seat of the stellar dynamo. It is also a key to understanding the $P_{rot}$--$P_{cyc}$ relation. The essential role of $55^\circ$ latitude in our dynamo model results from that the toroidal field is generated in the bulk of the convection zone by the latitudinal shear, rather than the widely assumed generation in the tachocline. In the bulk of the convection zone, the $55^\circ$ latitude has the strongest latitudinal shear. Although few dynamo models emphasize $55^\circ$ latitude, the importance of $55^\circ$ latitude in solar magnetism has been addressed by \cite{McIntosh2014,McIntosh2021} based on a wide variety of solar observations. The success of our model in reproducing the $P_{rot}$--$P_{cyc}$ relation of stellar activity adds a new piece of evidence on the important role of $55^\circ$ latitude in stellar magnetism, and further on the stellar dynamo working in the bulk of the convection zone. 

Being different from the $P_{rot}$--$P_{cyc}$ relation, the $P_{rot}$--$A_{cyc}$ relation did not pose a challenge for the past attempts to reproduce. For example, \cite{Karak2014, Kitchatinov2015} have explained the relation by taking into account the rotation-dependent tilt angle of starspots in their dynamos. Inspired by KN17 and observations relevant to gyrochronology \citep{Rengarajan1984,vanSaders2016}, we assume that faster-rotating stars operate in a more supercritical regime. The value $\alpha_0-\alpha_c$, describing how supercritical the dynamo is, is larger for faster rotators. The critical value of the BL source term, $\alpha_c$, corresponding to the marginal dynamo excitation, is calculated based on the linear models. The larger $\alpha_0-\alpha_c$ for faster rotators leads to a stronger saturated magnetic field, which explains the $P_{rot}$--$A_{cyc}$ relation in the nonlinear regime. The large supercriticality corresponding to the strong non-linearity for very fast rotators, e.g., $P_{rot}$ = 10 days, could entail fields with mixed parity. Higher latitude starspots for faster rotators make the coupling of poloidal flux between the two hemispheres more difficult so that the dominated parity for faster rotators could be quadrupolar shifted from the dipolar one for slower rotators. \cite{Hazra2019} have already reported a change in magnetic field parity from dipolar to quadrupolar in rapidly rotating stars. We wait for future stellar magnetic field measurement to verify the shift of magnetic field parity from mixed to quadrupolar, and further to dipolar when a star spins down with age because of the angular momentum loss.

\begin{acknowledgements}
Z.Z. and J.J. acknowledge financial support from the National Natural Science Foundation of China through grant Nos. 12173005 and 12350004, and the National Key	R$\&$D Program of China No. 2022YFF0503800. L.K. acknowledges financial support from the Ministry of Science and High Education of the Russian Federation.
\end{acknowledgements}

   \bibliographystyle{aa} 
   \bibliography{sample} 

\begin{thebibliography}{104}
\expandafter\ifx\csname natexlab\endcsname\relax\def\natexlab#1{#1}\fi

\bibitem[{{Augustson} {et~al.}(2012){Augustson}, {Brown}, {Brun}, {Miesch}, \&
  {Toomre}}]{Augustson2012}
{Augustson}, K.~C., {Brown}, B.~P., {Brun}, A.~S., {Miesch}, M.~S., \&
  {Toomre}, J. 2012, \apj, 756, 169

\bibitem[{{Augustson} {et~al.}(2019){Augustson}, {Brun}, \&
  {Toomre}}]{Augustson2019}
{Augustson}, K.~C., {Brun}, A.~S., \& {Toomre}, J. 2019, \apj, 876, 83

\bibitem[{{Baliunas} {et~al.}(1995){Baliunas}, {Donahue}, {Soon}, {Horne},
  {Frazer}, {Woodard-Eklund}, {Bradford}, {Rao}, {Wilson}, {Zhang}, {Bennett},
  {Briggs}, {Carroll}, {Duncan}, {Figueroa}, {Lanning}, {Misch}, {Mueller},
  {Noyes}, {Poppe}, {Porter}, {Robinson}, {Russell}, {Shelton}, {Soyumer},
  {Vaughan}, \& {Whitney}}]{Baliunas1995}
{Baliunas}, S.~L., {Donahue}, R.~A., {Soon}, W.~H., {et~al.} 1995, \apj, 438,
  269

\bibitem[{{Balona} \& {Abedigamba}(2016)}]{Balona2016}
{Balona}, L.~A. \& {Abedigamba}, O.~P. 2016, \mnras, 461, 497

\bibitem[{{Barnes} {et~al.}(2005){Barnes}, {Collier Cameron}, {Donati},
  {James}, {Marsden}, \& {Petit}}]{Barnes2005}
{Barnes}, J.~R., {Collier Cameron}, A., {Donati}, J.~F., {et~al.} 2005, \mnras,
  357, L1

\bibitem[{{Baumann} {et~al.}(2004){Baumann}, {Schmitt}, {Sch{\"u}ssler}, \&
  {Solanki}}]{Baumann2004}
{Baumann}, I., {Schmitt}, D., {Sch{\"u}ssler}, M., \& {Solanki}, S.~K. 2004,
  \aap, 426, 1075

\bibitem[{{B{\"o}hm-Vitense}(2007)}]{BohmVitense2007}
{B{\"o}hm-Vitense}, E. 2007, \apj, 657, 486

\bibitem[{{Bonanno} {et~al.}(2002){Bonanno}, {Elstner}, {R{\"u}diger}, \&
  {Belvedere}}]{Bonanno2002}
{Bonanno}, A., {Elstner}, D., {R{\"u}diger}, G., \& {Belvedere}, G. 2002, \aap,
  390, 673

\bibitem[{{Boro Saikia} {et~al.}(2018){Boro Saikia}, {Marvin}, {Jeffers},
  {Reiners}, {Cameron}, {Marsden}, {Petit}, {Warnecke}, \&
  {Yadav}}]{Saikia2018}
{Boro Saikia}, S., {Marvin}, C.~J., {Jeffers}, S.~V., {et~al.} 2018, \aap, 616,
  A108

\bibitem[{{Brown} {et~al.}(2008){Brown}, {Browning}, {Brun}, {Miesch}, \&
  {Toomre}}]{Brown2008}
{Brown}, B.~P., {Browning}, M.~K., {Brun}, A.~S., {Miesch}, M.~S., \& {Toomre},
  J. 2008, \apj, 689, 1354

\bibitem[{{Brun} \& {Browning}(2017)}]{Brun2017b}
{Brun}, A.~S. \& {Browning}, M.~K. 2017, Living Reviews in Solar Physics, 14, 4

\bibitem[{{Brun} {et~al.}(2015){Brun}, {Garc{\'\i}a}, {Houdek}, {Nandy}, \&
  {Pinsonneault}}]{Brun2015}
{Brun}, A.~S., {Garc{\'\i}a}, R.~A., {Houdek}, G., {Nandy}, D., \&
  {Pinsonneault}, M. 2015, \ssr, 196, 303

\bibitem[{{Brun} {et~al.}(2022){Brun}, {Strugarek}, {Noraz}, {Perri}, {Varela},
  {Augustson}, {Charbonneau}, \& {Toomre}}]{Brun2022}
{Brun}, A.~S., {Strugarek}, A., {Noraz}, Q., {et~al.} 2022, \apj, 926, 21

\bibitem[{{Caligari} {et~al.}(1995){Caligari}, {Moreno-Insertis}, \&
  {Schussler}}]{Caligari1995}
{Caligari}, P., {Moreno-Insertis}, F., \& {Schussler}, M. 1995, \apj, 441, 886

\bibitem[{{Cameron} \& {Sch{\"u}ssler}(2015)}]{Cameron2015}
{Cameron}, R. \& {Sch{\"u}ssler}, M. 2015, Science, 347, 1333

\bibitem[{{Cameron} {et~al.}(2010){Cameron}, {Jiang}, {Schmitt}, \&
  {Sch{\"u}ssler}}]{Cameron2010}
{Cameron}, R.~H., {Jiang}, J., {Schmitt}, D., \& {Sch{\"u}ssler}, M. 2010,
  \apj, 719, 264

\bibitem[{{Cameron} {et~al.}(2012){Cameron}, {Schmitt}, {Jiang}, \&
  {I{\c{s}}{\i}k}}]{Cameron2012}
{Cameron}, R.~H., {Schmitt}, D., {Jiang}, J., \& {I{\c{s}}{\i}k}, E. 2012,
  \aap, 542, A127

\bibitem[{{Cameron} \& {Sch{\"u}ssler}(2017)}]{Cameron2017}
{Cameron}, R.~H. \& {Sch{\"u}ssler}, M. 2017, \aap, 599, A52

\bibitem[{{Charbonneau}(2020)}]{Charbonneau2020}
{Charbonneau}, P. 2020, Living Reviews in Solar Physics, 17, 4

\bibitem[{{Chatterjee} {et~al.}(2004){Chatterjee}, {Nandy}, \&
  {Choudhuri}}]{Chatterjee2004}
{Chatterjee}, P., {Nandy}, D., \& {Choudhuri}, A.~R. 2004, \aap, 427, 1019

\bibitem[{{Choudhuri} {et~al.}(1995){Choudhuri}, {Schussler}, \&
  {Dikpati}}]{Choudhuri1995}
{Choudhuri}, A.~R., {Schussler}, M., \& {Dikpati}, M. 1995, \aap, 303, L29

\bibitem[{{Collier Cameron} \& {Donati}(2002)}]{CollierCameron2002}
{Collier Cameron}, A. \& {Donati}, J.~F. 2002, \mnras, 329, L23

\bibitem[{{Dasi-Espuig} {et~al.}(2010){Dasi-Espuig}, {Solanki}, {Krivova},
  {Cameron}, \& {Pe{\~n}uela}}]{DasiEspuig2010}
{Dasi-Espuig}, M., {Solanki}, S.~K., {Krivova}, N.~A., {Cameron}, R., \&
  {Pe{\~n}uela}, T. 2010, \aap, 518, A7

\bibitem[{{Dikpati} \& {Charbonneau}(1999)}]{Dikpati1999}
{Dikpati}, M. \& {Charbonneau}, P. 1999, \apj, 518, 508

\bibitem[{{Do Cao} \& {Brun}(2011)}]{DoCao2011}
{Do Cao}, O. \& {Brun}, A.~S. 2011, Astronomische Nachrichten, 332, 907

\bibitem[{{Durney}(1995)}]{Durney1995}
{Durney}, B.~R. 1995, \solphys, 160, 213

\bibitem[{{Fan}(2021)}]{Fan2021}
{Fan}, Y. 2021, Living Reviews in Solar Physics, 18, 5

\bibitem[{{G{\"u}del}(2004)}]{Gudel2004}
{G{\"u}del}, M. 2004, \aapr, 12, 71

\bibitem[{{Guerrero} \& {de Gouveia Dal Pino}(2008)}]{Guerrero2008}
{Guerrero}, G. \& {de Gouveia Dal Pino}, E.~M. 2008, \aap, 485, 267

\bibitem[{{Guerrero} {et~al.}(2019){Guerrero}, {Zaire}, {Smolarkiewicz}, {de
  Gouveia Dal Pino}, {Kosovichev}, \& {Mansour}}]{Guerrero2019}
{Guerrero}, G., {Zaire}, B., {Smolarkiewicz}, P.~K., {et~al.} 2019, \apj, 880,
  6

\bibitem[{{Hathaway}(2015)}]{Hathaway2015}
{Hathaway}, D.~H. 2015, Living Reviews in Solar Physics, 12, 4

\bibitem[{{Hazra} {et~al.}(2019){Hazra}, {Jiang}, {Karak}, \&
  {Kitchatinov}}]{Hazra2019}
{Hazra}, G., {Jiang}, J., {Karak}, B.~B., \& {Kitchatinov}, L. 2019, \apj, 884,
  35

\bibitem[{{Hazra} {et~al.}(2014){Hazra}, {Karak}, \& {Choudhuri}}]{Hazra2014}
{Hazra}, G., {Karak}, B.~B., \& {Choudhuri}, A.~R. 2014, \apj, 782, 93

\bibitem[{{Hazra} \& {Nandy}(2016)}]{Hazra2016}
{Hazra}, S. \& {Nandy}, D. 2016, \apj, 832, 9

\bibitem[{{Hempelmann} {et~al.}(1996){Hempelmann}, {Schmitt}, \&
  {St{\c{e}}pie{\'n}}}]{Hempelmann1996}
{Hempelmann}, A., {Schmitt}, J.~H.~M.~M., \& {St{\c{e}}pie{\'n}}, K. 1996,
  \aap, 305, 284

\bibitem[{{Hotta} \& {Yokoyama}(2010)}]{Hotta2010}
{Hotta}, H. \& {Yokoyama}, T. 2010, \apjl, 714, L308

\bibitem[{{I{\c{s}}{\i}k} {et~al.}(2018){I{\c{s}}{\i}k}, {Solanki}, {Krivova},
  \& {Shapiro}}]{Isik2018}
{I{\c{s}}{\i}k}, E., {Solanki}, S.~K., {Krivova}, N.~A., \& {Shapiro}, A.~I.
  2018, \aap, 620, A177

\bibitem[{{Jennings} \& {Weiss}(1991)}]{Jennings1991}
{Jennings}, R.~L. \& {Weiss}, N.~O. 1991, \mnras, 252, 249

\bibitem[{{Jiang}(2020)}]{Jiang2020}
{Jiang}, J. 2020, \apj, 900, 19

\bibitem[{{Jiang} {et~al.}(2013){Jiang}, {Cameron}, {Schmitt}, \&
  {I{\c{s}}{\i}k}}]{Jiang2013}
{Jiang}, J., {Cameron}, R.~H., {Schmitt}, D., \& {I{\c{s}}{\i}k}, E. 2013,
  \aap, 553, A128

\bibitem[{{Jiang} {et~al.}(2011){Jiang}, {Cameron}, {Schmitt}, \&
  {Sch{\"u}ssler}}]{Jiang2011}
{Jiang}, J., {Cameron}, R.~H., {Schmitt}, D., \& {Sch{\"u}ssler}, M. 2011,
  \aap, 528, A82

\bibitem[{{Jiang} {et~al.}(2014{\natexlab{a}}){Jiang}, {Cameron}, \&
  {Sch{\"u}ssler}}]{Jiang2014}
{Jiang}, J., {Cameron}, R.~H., \& {Sch{\"u}ssler}, M. 2014{\natexlab{a}}, \apj,
  791, 5

\bibitem[{{Jiang} {et~al.}(2014{\natexlab{b}}){Jiang}, {Hathaway}, {Cameron},
  {Solanki}, {Gizon}, \& {Upton}}]{Jiang2014b}
{Jiang}, J., {Hathaway}, D.~H., {Cameron}, R.~H., {et~al.} 2014{\natexlab{b}},
  \ssr, 186, 491

\bibitem[{{Jiang} \& {Wang}(2006)}]{Jiang2005}
{Jiang}, J. \& {Wang}, J.-X. 2006, \cjaa, 6, 227

\bibitem[{{Jiang} \& {Wang}(2007)}]{Jiang2007}
{Jiang}, J. \& {Wang}, J.~X. 2007, \mnras, 377, 711

\bibitem[{{Jiang} {et~al.}(2018){Jiang}, {Wang}, {Jiao}, \& {Cao}}]{Jiang2018}
{Jiang}, J., {Wang}, J.-X., {Jiao}, Q.-R., \& {Cao}, J.-B. 2018, \apj, 863, 159

\bibitem[{{Jiao} {et~al.}(2021){Jiao}, {Jiang}, \& {Wang}}]{Jiao2021}
{Jiao}, Q., {Jiang}, J., \& {Wang}, Z.-F. 2021, \aap, 653, A27

\bibitem[{{Jouve} {et~al.}(2010){Jouve}, {Brown}, \& {Brun}}]{Jouve2010}
{Jouve}, L., {Brown}, B.~P., \& {Brun}, A.~S. 2010, \aap, 509, A32

\bibitem[{{Jouve} \& {Brun}(2007)}]{Jouve2007}
{Jouve}, L. \& {Brun}, A.~S. 2007, \aap, 474, 239

\bibitem[{{Jouve} {et~al.}(2008){Jouve}, {Brun}, {Arlt}, {Brandenburg},
  {Dikpati}, {Bonanno}, {K{\"a}pyl{\"a}}, {Moss}, {Rempel}, {Gilman}, {Korpi},
  \& {Kosovichev}}]{Jouve2008}
{Jouve}, L., {Brun}, A.~S., {Arlt}, R., {et~al.} 2008, \aap, 483, 949

\bibitem[{{Karak}(2010)}]{Karak2010}
{Karak}, B.~B. 2010, \apj, 724, 1021

\bibitem[{{Karak}(2020)}]{Karak2020}
{Karak}, B.~B. 2020, \apjl, 901, L35

\bibitem[{{Karak} {et~al.}(2014{\natexlab{a}}){Karak}, {Jiang}, {Miesch},
  {Charbonneau}, \& {Choudhuri}}]{Karak2014b}
{Karak}, B.~B., {Jiang}, J., {Miesch}, M.~S., {Charbonneau}, P., \&
  {Choudhuri}, A.~R. 2014{\natexlab{a}}, \ssr, 186, 561

\bibitem[{{Karak} {et~al.}(2014{\natexlab{b}}){Karak}, {Kitchatinov}, \&
  {Choudhuri}}]{Karak2014}
{Karak}, B.~B., {Kitchatinov}, L.~L., \& {Choudhuri}, A.~R. 2014{\natexlab{b}},
  \apj, 791, 59

\bibitem[{{Kitchatinov}(2022)}]{Kitchatinov2022}
{Kitchatinov}, L. 2022, Research in Astronomy and Astrophysics, 22, 125006

\bibitem[{{Kitchatinov} \&
  {Nepomnyashchikh}(2017{\natexlab{a}})}]{Kitchatinov2017a}
{Kitchatinov}, L. \& {Nepomnyashchikh}, A. 2017{\natexlab{a}}, \mnras, 470,
  3124

\bibitem[{{Kitchatinov} \&
  {Nepomnyashchikh}(2017{\natexlab{b}})}]{Kitchatinov2017b}
{Kitchatinov}, L.~L. \& {Nepomnyashchikh}, A.~A. 2017{\natexlab{b}}, Astronomy
  Letters, 43, 332

\bibitem[{{Kitchatinov} \& {Olemskoy}(2011)}]{Kitchatinov2011}
{Kitchatinov}, L.~L. \& {Olemskoy}, S.~V. 2011, \mnras, 411, 1059

\bibitem[{{Kitchatinov} \& {Olemskoy}(2012)}]{Kitchatinov2012}
{Kitchatinov}, L.~L. \& {Olemskoy}, S.~V. 2012, \mnras, 423, 3344

\bibitem[{{Kitchatinov} \& {Olemskoy}(2015)}]{Kitchatinov2015}
{Kitchatinov}, L.~L. \& {Olemskoy}, S.~V. 2015, Research in Astronomy and
  Astrophysics, 15, 1801

\bibitem[{{Kitchatinov} \& {R{\"u}diger}(1999)}]{Kitchatinov1999}
{Kitchatinov}, L.~L. \& {R{\"u}diger}, G. 1999, \aap, 344, 911

\bibitem[{{Kochukhov}(2021)}]{Kochukhov2021}
{Kochukhov}, O. 2021, \aapr, 29, 1

\bibitem[{{Li} {et~al.}(2003){Li}, {Wang}, {Zhan}, {Yun}, {Liang}, {Zhao}, \&
  {Gu}}]{Li2003}
{Li}, K.~J., {Wang}, J.~X., {Zhan}, L.~S., {et~al.} 2003, \solphys, 215, 99

\bibitem[{{McIntosh} {et~al.}(2021){McIntosh}, {Leamon}, {Egeland}, {Dikpati},
  {Altrock}, {Banerjee}, {Chatterjee}, {Srivastava}, \& {Velli}}]{McIntosh2021}
{McIntosh}, S.~W., {Leamon}, R.~J., {Egeland}, R., {et~al.} 2021, \solphys,
  296, 189

\bibitem[{{McIntosh} {et~al.}(2014){McIntosh}, {Wang}, {Leamon}, {Davey},
  {Howe}, {Krista}, {Malanushenko}, {Markel}, {Cirtain}, {Gurman}, {Pesnell},
  \& {Thompson}}]{McIntosh2014}
{McIntosh}, S.~W., {Wang}, X., {Leamon}, R.~J., {et~al.} 2014, \apj, 792, 12

\bibitem[{{Montet} {et~al.}(2017){Montet}, {Tovar}, \&
  {Foreman-Mackey}}]{Montet2017}
{Montet}, B.~T., {Tovar}, G., \& {Foreman-Mackey}, D. 2017, \apj, 851, 116

\bibitem[{{Mu{\~n}oz-Jaramillo} {et~al.}(2011){Mu{\~n}oz-Jaramillo}, {Nandy},
  \& {Martens}}]{Munoz2011}
{Mu{\~n}oz-Jaramillo}, A., {Nandy}, D., \& {Martens}, P. C.~H. 2011, \apjl,
  727, L23

\bibitem[{Mu{\~n}oz-Jaramillo {et~al.}(2013)Mu{\~n}oz-Jaramillo, Dasi-Espuig,
  Balmaceda, \& DeLuca}]{Jaramillo2013}
Mu{\~n}oz-Jaramillo, A., Dasi-Espuig, M., Balmaceda, L.~A., \& DeLuca, E.~E.
  2013, The Astrophysical Journal Letters, 767, L25

\bibitem[{{Nandy}(2004)}]{Nandy2004}
{Nandy}, D. 2004, \solphys, 224, 161

\bibitem[{{Nelson} {et~al.}(2013){Nelson}, {Brown}, {Brun}, {Miesch}, \&
  {Toomre}}]{Nelson2013}
{Nelson}, N.~J., {Brown}, B.~P., {Brun}, A.~S., {Miesch}, M.~S., \& {Toomre},
  J. 2013, \apj, 762, 73

\bibitem[{{N{\`e}mec} {et~al.}(2023){N{\`e}mec}, {Shapiro}, {I{\c{s}}{\i}k},
  {Solanki}, \& {Reinhold}}]{Nemec2023}
{N{\`e}mec}, N.~E., {Shapiro}, A.~I., {I{\c{s}}{\i}k}, E., {Solanki}, S.~K., \&
  {Reinhold}, T. 2023, \aap, 672, A138

\bibitem[{{Noyes} {et~al.}(1984{\natexlab{a}}){Noyes}, {Hartmann}, {Baliunas},
  {Duncan}, \& {Vaughan}}]{Noyes1984a}
{Noyes}, R.~W., {Hartmann}, L.~W., {Baliunas}, S.~L., {Duncan}, D.~K., \&
  {Vaughan}, A.~H. 1984{\natexlab{a}}, \apj, 279, 763

\bibitem[{{Noyes} {et~al.}(1984{\natexlab{b}}){Noyes}, {Weiss}, \&
  {Vaughan}}]{Noyes1984b}
{Noyes}, R.~W., {Weiss}, N.~O., \& {Vaughan}, A.~H. 1984{\natexlab{b}}, \apj,
  287, 769

\bibitem[{{Parker}(1955)}]{Parker1955b}
{Parker}, E.~N. 1955, \apj, 121, 491

\bibitem[{{Petrovay} {et~al.}(2020){Petrovay}, {Nagy}, \&
  {Yeates}}]{Petrovay_2020}
{Petrovay}, K., {Nagy}, M., \& {Yeates}, A.~R. 2020, Journal of Space Weather
  and Space Climate, 10, 50

\bibitem[{{Pipin} \& {Kosovichev}(2016)}]{Pipin2016}
{Pipin}, V.~V. \& {Kosovichev}, A.~G. 2016, \apj, 823, 133

\bibitem[{{Reinhold} {et~al.}(2017){Reinhold}, {Cameron}, \&
  {Gizon}}]{Reinhold2017}
{Reinhold}, T., {Cameron}, R.~H., \& {Gizon}, L. 2017, \aap, 603, A52

\bibitem[{{Rengarajan}(1984)}]{Rengarajan1984}
{Rengarajan}, T.~N. 1984, \apjl, 283, L63

\bibitem[{{Saar} \& {Brandenburg}(1999)}]{Saar1999}
{Saar}, S.~H. \& {Brandenburg}, A. 1999, \apj, 524, 295

\bibitem[{{Schou} {et~al.}(1998){Schou}, {Antia}, {Basu}, {Bogart}, {Bush},
  {Chitre}, {Christensen-Dalsgaard}, {Di Mauro}, {Dziembowski}, {Eff-Darwich},
  {Gough}, {Haber}, {Hoeksema}, {Howe}, {Korzennik}, {Kosovichev}, {Larsen},
  {Pijpers}, {Scherrer}, {Sekii}, {Tarbell}, {Title}, {Thompson}, \&
  {Toomre}}]{Schou1998}
{Schou}, J., {Antia}, H.~M., {Basu}, S., {et~al.} 1998, \apj, 505, 390

\bibitem[{{Schrijver} {et~al.}(1996){Schrijver}, {Shine}, {Hagenaar},
  {Hurlburt}, {Title}, {Strous}, {Jefferies}, {Jones}, {Harvey}, \&
  {Duvall}}]{Schrijver1996}
{Schrijver}, C.~J., {Shine}, R.~A., {Hagenaar}, H.~J., {et~al.} 1996, \apj,
  468, 921

\bibitem[{{Schrijver} \& {Title}(2001)}]{Schrijver2001}
{Schrijver}, C.~J. \& {Title}, A.~M. 2001, \apj, 551, 1099

\bibitem[{{Schuessler} \& {Solanki}(1992)}]{Schuessler1992}
{Schuessler}, M. \& {Solanki}, S.~K. 1992, \aap, 264, L13

\bibitem[{{Solanki} {et~al.}(2008){Solanki}, {Wenzler}, \&
  {Schmitt}}]{Solanki2008}
{Solanki}, S.~K., {Wenzler}, T., \& {Schmitt}, D. 2008, \aap, 483, 623

\bibitem[{{Sowmya} {et~al.}(2021){Sowmya}, {N{\`e}mec}, {Shapiro},
  {I{\c{s}}{\i}k}, {Witzke}, {Mints}, {Krivova}, \& {Solanki}}]{Sowmya2021}
{Sowmya}, K., {N{\`e}mec}, N.~E., {Shapiro}, A.~I., {et~al.} 2021, \apj, 919,
  94

\bibitem[{{Strassmeier}(2009)}]{Strassmeier2009}
{Strassmeier}, K.~G. 2009, \aapr, 17, 251

\bibitem[{{Strassmeier} {et~al.}(1991){Strassmeier}, {Rice}, {Wehlau}, {Vogt},
  {Hatzes}, {Tuominen}, {Piskunov}, {Hackman}, \& {Poutanen}}]{Strassmeier1991}
{Strassmeier}, K.~G., {Rice}, J.~B., {Wehlau}, W.~H., {et~al.} 1991, \aap, 247,
  130

\bibitem[{{Strugarek} {et~al.}(2017){Strugarek}, {Beaudoin}, {Charbonneau},
  {Brun}, \& {do Nascimento}}]{Strugarek2017}
{Strugarek}, A., {Beaudoin}, P., {Charbonneau}, P., {Brun}, A.~S., \& {do
  Nascimento}, J.~D. 2017, Science, 357, 185

\bibitem[{{Talafha} {et~al.}(2022){Talafha}, {Nagy}, {Lemerle}, \&
  {Petrovay}}]{Talafha2022}
{Talafha}, M., {Nagy}, M., {Lemerle}, A., \& {Petrovay}, K. 2022, \aap, 660,
  A92

\bibitem[{{Tobias}(1998)}]{Tobias1998}
{Tobias}, S.~M. 1998, \mnras, 296, 653

\bibitem[{{van der Houwen} \& {Sommeijer}(2001)}]{Houwen2001}
{van der Houwen}, P.~J. \& {Sommeijer}, B.~P. 2001, Journal of Computational
  and Applied Mathematics, 128, 447

\bibitem[{{van Saders} {et~al.}(2016){van Saders}, {Ceillier}, {Metcalfe},
  {Silva Aguirre}, {Pinsonneault}, {Garc{\'\i}a}, {Mathur}, \&
  {Davies}}]{vanSaders2016}
{van Saders}, J.~L., {Ceillier}, T., {Metcalfe}, T.~S., {et~al.} 2016, \nat,
  529, 181

\bibitem[{{Vashishth} {et~al.}(2023){Vashishth}, {Karak}, \&
  {Kitchatinov}}]{Vashishth2023}
{Vashishth}, V., {Karak}, B.~B., \& {Kitchatinov}, L. 2023, \mnras, 522, 2601

\bibitem[{{Vogt} \& {Penrod}(1983)}]{Vogt1983}
{Vogt}, S.~S. \& {Penrod}, G.~D. 1983, \pasp, 95, 565

\bibitem[{{Wang} {et~al.}(1991){Wang}, {Sheeley}, \& {Nash}}]{Wang1991}
{Wang}, Y.~M., {Sheeley}, N.~R., J., \& {Nash}, A.~G. 1991, \apj, 383, 431

\bibitem[{{Wang} \& {Sheeley}(2002)}]{Wang2001}
{Wang}, Y.~M. \& {Sheeley}, N.~R. 2002, Journal of Geophysical Research (Space
  Physics), 107, 1302

\bibitem[{{Warnecke}(2018)}]{Warnecke2018b}
{Warnecke}, J. 2018, \aap, 616, A72

\bibitem[{{Wilson}(1978)}]{Wilson1978}
{Wilson}, O.~C. 1978, \apj, 226, 379

\bibitem[{{Wright} \& {Drake}(2016)}]{Wright2016}
{Wright}, N.~J. \& {Drake}, J.~J. 2016, \nat, 535, 526

\bibitem[{{Wright} {et~al.}(2011){Wright}, {Drake}, {Mamajek}, \&
  {Henry}}]{Wright2011}
{Wright}, N.~J., {Drake}, J.~J., {Mamajek}, E.~E., \& {Henry}, G.~W. 2011,
  \apj, 743, 48

\bibitem[{{Yadav} {et~al.}(2015){Yadav}, {Gastine}, {Christensen}, \&
  {Reiners}}]{Yadav2015}
{Yadav}, R.~K., {Gastine}, T., {Christensen}, U.~R., \& {Reiners}, A. 2015,
  \aap, 573, A68

\bibitem[{{Zhang} {et~al.}(2020){Zhang}, {Bi}, {Li}, {Jiang}, {Li}, {He}, {Yu},
  {Khanna}, {Ge}, {Liu}, {Tian}, {Wu}, \& {Zhang}}]{Zhang2020}
{Zhang}, J., {Bi}, S., {Li}, Y., {et~al.} 2020, \apjs, 247, 9

\bibitem[{{Zhang} \& {Jiang}(2022)}]{Zhang2022a}
{Zhang}, Z. \& {Jiang}, J. 2022, \apj, 930, 30

\bibitem[{{Zhang} {et~al.}(2022){Zhang}, {Jiang}, \& {Zhang}}]{Zhang2022b}
{Zhang}, Z., {Jiang}, J., \& {Zhang}, H. 2022, \apjl, 941, L3

\end{thebibliography}

\end{document}